\documentclass[pra,twocolumn,tbtags,superscriptaddress,floatfix,showpacs]{revtex4}
\usepackage{amssymb}
\usepackage{graphicx,amsmath}
\usepackage{bm}
\usepackage{times}
\usepackage{float}

\begin{document}
\title{Mollow triplet: pump probe single photon spectroscopy of artificial atoms}

\begin{abstract}
We analyze a photon transport through an 1D open waveguide side
coupled to the $N$-photon microwave cavity with embedded
artificial two- level atom (qubit). The qubit state is probed by a
weak signal at the fundamental frequency of the waveguide. Within
the formalism of projection operators and non-Hermitian
Hamiltonian approach we develop a one-photon approximation scheme
to obtain the photon wavefunction which allows for the calculation
of the probability amplitudes of the spontaneous transitions
between the levels of two Rabi doublets in $N$- photon cavity. We
obtain analytic expressions for the transmission and reflection
factors of the microwave signal through a waveguide, which contain
the information of the qubit parameters. We show that for small
number of cavity photons the Mollow spectrum consists of four
spectral lines which is a direct manifestation of quantum nature
of light. The results obtained in the paper are of general nature
and can be applied to any type of qubits. The specific properties
of the qubit are only encoded in the two parameters: the  energy
$\Omega$ of the qubit and its coupling $\lambda$ to the cavity
photons.


\end{abstract}


\date{\today }
\author{Ya. S. Greenberg}\email{greenbergy@risp.ru}
\affiliation{Novosibirsk State Technical University, Novosibirsk,
Russia}

\author{A. N. Sultanov}\email{sultanov.aydar@ngs.ru}
\affiliation{Novosibirsk State Technical University, Novosibirsk,
Russia}


 \maketitle
 \section{Introduction}\label{Intr}

The coherent coupling of a superconducting qubit to the microwave
modes of a 1D coplanar waveguide transmission line has been
intensely investigated over the last years  both experimentally
and theoretically. As compared with the conventional optical
cavity with atomic gases, superconducting qubits as artificial
atoms in solid-state devices have significant advantages, such as
technological scalability, long coherence time which is important
for the implementation of the quantum gate operations, huge
tunability and controllability by external electromagnetic fields
\cite{You11, Girvin09, Pashkin09, Sanders11}. Another advantage is
an on-chip realization of strong and ultrastrong coupling regimes
\cite{Wal04, Niemczyk10} previously inaccessible to atomic
systems. This enables us to explore novel quantum phenomena
emerging only in this regime. Furthermore, solid state
superconducting circuits with embedded Josephson junction qubits
have reproduced many physical phenomena known previously from
quantum optics, such as Kerr nonlinearities \cite{Rebic09,
Rehak14}, electromagnetically induced transparency \cite{Sun14,
Abdumalikov10, Joo10, Li13}, the Mollow triplet \cite{Baur09,
Astafiev10, Hoi12, Lang11, Toyli16}, and Autler-Townes splitting
\cite{Sun14, Baur09, Sillanpaa09}.

As the Mollow triplet is a clear manifestation of the coherent
nature of the light- matter interaction, its fluorescent or
transmission spectra can be explained considering the pumping
light classically \cite{Mollow69}. Instead of looking at the
emission fluorescent spectrum we study here the transmission of a
single photon which induces the transitions in a preliminary
pumped cavity. The use of a single photon source as a probe
reveals a marked influence of the quantum nature of light on the
Mollow spectra  and allows us to determine the response to the
input of a single injected photon \cite{Lang11, Toyli16, Fink08}.
Thus, a theoretical framework that allows one to directly
calculate the response of such a system to a single injected
photon is justified.

A conventional technique, which is used to study the photon
transport in 1D geometry, is based on the master equation for the
density matrix. It allows one to find analytic solution only for
$N=1$ \cite{Om10}. For $N>1$ the solution of master equation are
usually being solved approximately by numerical integration
\cite{Bian09}. As to our knowledge, even for $N=2$ the analytic
expressions for photon transport coefficients are not known.

From the other point, this technique is not quite suitable for
single photon measurements since it operates with the average
quantities. More appropriate approach for the description of a
single photon transport is the calculation of the photon
wavefunction which carries the information about quantum dynamics
of the photon- matter interaction \cite{Shen05, Shen09}.

In the present paper we consider the transmission and reflection
Mollow spectra for artificial atom (qubit) embedded in the $N$-
photon cavity which is side- coupled to open microwave waveguide.
We find the explicit expressions for the photon wavefunctions
which describe the scattering of a single photon on the
atom-cavity system with any value $N$ of preliminary pumped cavity
photons.

Our analysis is based on the projection operators formalism and
the method of the effective non- Hermitian Hamiltonian which has
many applications for different open mesoscopic systems (see
review paper \cite{Auerbach11} and references therein). Recently
this method has been applied to photon transport through 1D open
transmission line with $N$ embedded qubits \cite{Greenberg15}.

We find the analytic expressions for the probability amplitudes of
the spontaneous transitions induced by injected photon in $N$-
photon cavity. This enables us to find the forms of spectral lines
depending on the qubit parameters and on the number of photons in
a cavity. We show that for small number of cavity photons the
transmission and reflection spectra consist of four lines which is
a direct manifestation of quantum nature of light. As the number
of cavity photons is increased two central peaks merge giving a
conventional Mollow triplet.

The results obtained in the paper are relevant for the experiments
where a qubit+cavity system is preliminary being driven by a
fixed-frequency pump field to one of its excited $N$- photon
states, with transitions to higher-lying states being studied by a
weak, variable-frequency single photon probe \cite{Fink08}.
Another application of our results is a phenomenon which is called
a photon blockade. The excitation of the nonlinear atom-cavity
system by a first photon at the frequency $\omega$ blocks the
transmission of a second photon at the same frequency \cite
{Birnbaum05}.

The paper is organized as follows. In Sec.\ref{Proj} we briefly
describe the projection operators formalism and the method of
effective non hermitian Hamiltonian. In Sec.\ref{Scat} we define
the Hamiltonian of 1D waveguide side coupled to the $N$- photon
microwave cavity with embedded qubit and qualitatively describe
the process of a single photon scattering. The analytical
expression for the effective non-hermitian Hamiltonian is given in
Sec.\ref{Cav}. In this section we find the spectrum of the cavity
resonances and their dependence on the cavity decay rate $\Gamma$,
cavity-qubit coupling strength $\lambda$ and the number of cavity
photons $N$. The wave function of the scattering photon is found
in Sec.\ref{MT}, where we obtain the explicit analytical
expressions for the probability amplitudes which describe
spontaneous transitions between the levels of two Rabi doublets.
These amplitudes are directly related to the transmission and
reflection factors and show representative photon spectra. The
results obtained in Sec.\ref{MT} are applied in Sec.\ref{TS} where
the transmission amplitudes have been analyzed in detail.  In
Sec.\ref{experFink} we analyzed the case $N=2$ and showed that our
results are consistent with the experiment in \cite{Fink08}. In
addition, we show in Sec.\ref{experFink} that in the experimental
scheme of Fink et. al. \cite{Fink08} our results predicts the
detection of a single photon with the frequency which is shifted
from that of the input photon by a Rabi frequency. The probability
amplitudes for this process are calculated. The application of our
results to the description of photon blockade is given in
Sec.\ref{phb}.

\section{Projection formalism and effective non-hermitian
Hamiltonian}\label{Proj}

We start  with a brief review of projection formalism,
highlighting only those aspects that will be required for the
paper here. The application of this method to photon transport was
described in more detail in \cite{Greenberg15}.

According to this method the Hilbert space of a quantum system
with the Hermitian Hamiltonian $H$ is formally subdivided  into
two arbitrarily selected orthogonal projectors, $P$ and $Q$, which
satisfy the following properties:

\begin{equation}\label{P+Q}
    P+Q=1;\quad PQ=QP=0; \quad PP=P ;\quad QQ=Q
\end{equation}

Keeping in mind the scattering problem we assume that $Q$ subspace
determines a closed system and, therefore, consists of discrete
states, and $P$ subspace consists of the states from continuum.
Those states of subspace Q which will turn out to be coupled to
the states in subspace P will acquire the outgoing waves and
become unstable. Then, for this scattering problem the effective
Hamiltonian which describes the decay of Q- subsystem, becomes non
Hermitian and has to be written as follows:


\begin{equation}\label{Heff1}
    {H_{eff}(E)} = {H_{QQ}} + {H_{QP}}\frac{1}{{E - {H_{PP}+i\varepsilon} }}{H_{PQ}}
\end{equation}
where $ H_{XY}=XHY$, with $X,Y$ being $Q$ or $P$.


The effective Hamiltonian (\ref{Heff1}) determines the resonance
energies of the Q- subsystem which are due to its interaction
$H_{PQ}$ with continuum states from P- system. These resonances
lie in the low half of the complex energy plane,
$z=\rm{\widetilde{E}}-i\hbar\widetilde{\Gamma}$ and are given by
the roots of the equation
\begin{equation}\label{2qb2}
    D(z)\equiv\det \left( {z - {H_{eff}}} \right) = 0
\end{equation}
The imaginary part $\widetilde{\Gamma}$ of the resonances
describes the decay of $Q$- states due to their interaction with
continuum $P$- states.

The scattering solution for the state vector of the Shr\"{o}dinger
equation $H\Psi=E\Psi$ reads \cite{Rotter01}

\begin{eqnarray}\label{Ph1}
    |{\Psi}\rangle  = \left| {in}\right\rangle
   + \frac{1}{{E - {H_{eff}}}}{H_{QP}}\left| {in} \right\rangle \nonumber\\
   +\frac{1}{{E - {H_{PP}} + i\varepsilon }} H_{PQ}\frac{1}{{E - {H_{eff}}}}{H_{QP}}\left| {in} \right\rangle\
\end{eqnarray}
where $|in\rangle$ is the initial state, which contains continuum
variables and satisfies the equation $H_{PP}|in \rangle =E |in
\rangle$.

The last term in the expression (\ref{Ph1}) describes to all
orders of $H_{QP}$ the evolution of initial state $|in\rangle$
under the interaction between $P$ and $Q$ subspaces.

It is useful to stress that the formal results (\ref{Heff1}) and
(\ref{Ph1}) do not require any explicit expressions for the
projection operators.

\section{Single photon scattering}\label{Scat}

\begin{figure}
  \includegraphics[width=8cm,keepaspectratio]{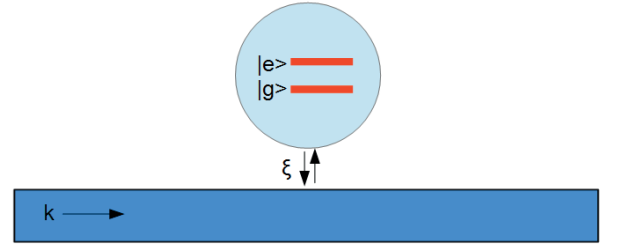}
  \caption{(Color online) Waveguide side coupled to the $N$- photon cavity with
  imbedded qubit}\label{SC}
\end{figure}

We consider a microwave 1D waveguide side coupled to a cavity with
embedded qubit as is shown in Fig.\ref{SC}.

The Hamiltonian of the system reads:
\begin{eqnarray}\label{H_ph}
H = \sum\limits_k {\hbar {\omega _k}c_k^ + {c_k}}
 + \frac{1}{2}\hbar \Omega {\sigma _z} + \hbar {\omega _c}{a^ + }a + \hbar \lambda (a_{}^ +
  + a){\sigma _X}\nonumber\\
 + \hbar \xi \sum\limits_k {(c_k^ + a + {c_k}{a^ + })}\quad\quad\quad\quad\quad\quad\quad\quad\quad\quad\quad\quad
\end{eqnarray}

where the first three terms are, respectively the Hamiltonian of
waveguide photons, the Hamiltonian of the  qubit with the
excitation frequency $\Omega$,  the Hamiltonian of one mode
cavity. Fourth and fifth terms describe the qubit-cavity
interaction with the strength $\lambda$, and the interaction
between the waveguide and the cavity with the strength $\xi$.

As we study a single photon probe we assume that at every instant
there is either one photon in a waveguide and $N-1$ photons in a
cavity or no photons in a waveguide and $N$ photons in a cavity.
Therefore, we assume that our Hilbert space is restricted to the
following state vectors:

\begin{equation}\label{Q}
\left| 1 \right\rangle  \equiv \left| 0 \right\rangle  \otimes
\left| {g,N} \right\rangle {\mkern 1mu} ,{\mkern 1mu} {\mkern 1mu}
\left| 2 \right\rangle  \equiv \left| 0 \right\rangle  \otimes
\left| {e,N - 1} \right\rangle
\end{equation}

\begin{equation}\label{P}
\left| {{k_1}} \right\rangle  \equiv \left| k \right\rangle
\otimes \left| {g,N - 1} \right\rangle {\mkern 1mu} ,{\mkern 1mu}
{\mkern 1mu} \left| {{k_2}} \right\rangle  \equiv \left| k
\right\rangle  \otimes \left| {e,N - 2} \right\rangle
\end{equation}

The states (\ref{Q}) correspond to no photons in a waveguide, $N$
photons in the cavity, and a qubit in the ground $g$ or excited
 state $e$. The states (\ref{P}) correspond to the situation where one photon
with a momentum $k$ is in a waveguide, $N-1$ photons in the
cavity, and a qubit in the ground $g$ or excited  state $e$.

Due to the interaction between cavity photons and a qubit each of
the pair of states (\ref{Q}) and  (\ref{P}) are being hybridized
to two pairs of dressed states
$|\chi_{i,0}\rangle=|0\rangle\otimes |\chi_i\rangle$,
$|\varphi_{i,k}\rangle=|k\rangle\otimes |\varphi_i\rangle$, where

\begin{equation}\label{chi}
    |\chi_i\rangle=\alpha_i|g,N\rangle +\beta_i|e,N-1\rangle
\end{equation}

\begin{equation}\label{fi}
    |\varphi_i\rangle=a_i|g,N-1\rangle+b_i|e,N-2\rangle
\end{equation}

Every pair of these dressed states are split by a Rabi frequency
corresponding to the number of the cavity photons:
\begin{equation}\label{rabi}
    \Omega_R^{(N)} = \sqrt{{\Delta^2} + 4{\lambda ^2}N}
\end{equation}

where $\Delta=\omega_c-\Omega$.

For subsequent calculations we need only the explicit form of the
superposition factors $a_i$ and $b_i$ in (\ref{fi}) which can be
expressed in terms of the angle variable $\theta$: $\tan
2\theta=-2\lambda(N-1)/\Delta$
 with $a_1=b_2=\sin\theta$, $b_1=-a_2=\cos\theta$.

\begin{figure}
  \includegraphics[width=8 cm]{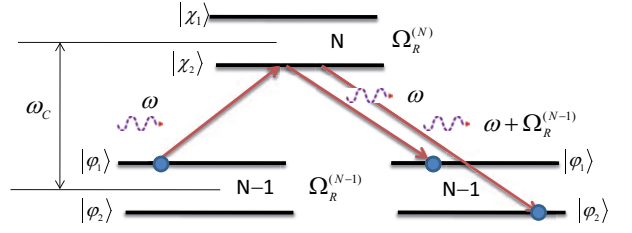}\\
  \caption{Color online. A scheme of a scattering of a single photon on the
state $|\varphi_1\rangle$ of $(N-1)$- cavity when the transition
$|\varphi_1\rangle\rightarrow|\chi_2\rangle$ is
excited.}\label{tr11}
\end{figure}

The process of the photon scattering can be qualitatively
described as follows. Before a probing photon enters a waveguide
the $N-1$ photon cavity + qubit system is in one of its hybridized
states $|\varphi_i\rangle$($i=1,2$) (\ref{fi}) that was prepared
by a preliminary pumping. The multiple interaction of a probing
photon with a cavity leads to the formation  of quasienergy
hybridized states (\ref{chi}). These states subsequently decay
with one photon being escaped to a waveguide, and  a cavity +
qubit system being left in one of the states (\ref{fi}). This
picture is illustrated in Fig. \ref{tr11} where the incoming
photon excites the $|\varphi_1\rangle$ state to the state
$|\chi_2\rangle$ at the frequency
$\omega=\omega_C-\frac{1}{2}(\Omega_R^{(N)}+\Omega_R^{(N-1)})$.
The state $|\chi_2\rangle$ subsequently decays either to the
initial state $|\varphi_1\rangle$ with the outgoing photon having
the excitation frequency $\omega$ or to the state
$|\varphi_2\rangle$ with the outgoing photon having the frequency
$\omega+\Omega_R^{(N-1)}$ .

\begin{figure}[h]
  \includegraphics[width=8 cm, keepaspectratio]{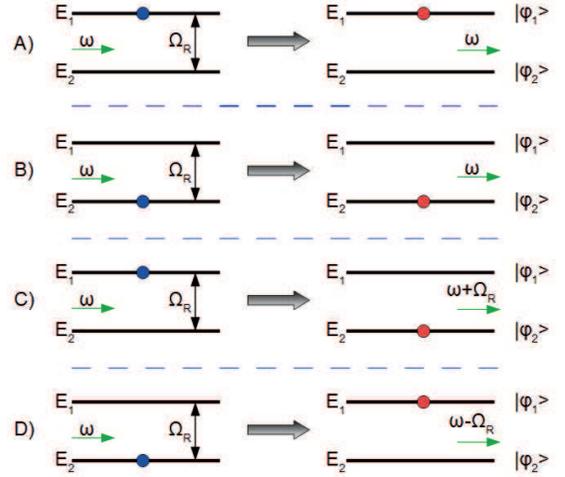}\\
  \caption{(Color online) Four outcomes of the scattering process.
  Two upper graphs correspond to elastic scattering, while two
  lower graphs correspond to inelastic sidebands.
   Blue circles denote the initial state,
   red ones denote the final state.}\label{Fig2}
\end{figure}

Hence there four possible  outcomes of a probing photon scattering
depending on which of the two states (\ref{fi}) were prepared by a
preliminary pumping. These four possible channels are shown in
Fig.\ref{Fig2}.

 Two channels describe the elastic scattering when
the initial and final states of the $N-1$ cavity + qubit system
before and after scattering are the same, and the energies of
incoming and outgoing photons are equal. The other two channels
describe inelastic process when the outgoing photon gains or loses
its energy by amount of $\hbar\Omega_R^{(N-1)}$. Every channel
shown in Fig.\ref{Fig2} corresponds to a specific transmission
factor that will be calculated below. Each channel has two
resonances which correspond to two transitions from $N$ photon
cavity to one of the final states $\varphi_i$. For example, the
channel $A$ in Fig.\ref{Fig2} has one resonance at the frequency
$\omega=\omega_C-\frac{1}{2}(\Omega_R^{(N)}+\Omega_R^{(N-1)})$
that induces the transition
$|\varphi_1\rangle\rightarrow|\chi_2\rangle$ (see Fig.\ref{tr11}).
The other resonance is at the frequency
$\omega=\omega_C+\frac{1}{2}(\Omega_R^{(N)}-\Omega_R^{(N-1)})$
that induces the transition
$|\varphi_1\rangle\rightarrow|\chi_1\rangle$. Each of these
resonances subsequently decays to the initial state
$|\varphi_1\rangle$.

\section{Cavity resonances}\label{Cav}

 In accordance with the projection operators formalism we define two
mutual orthogonal subspaces as follows

\begin{equation}\label{QQ}
    Q=|1\rangle\langle 1|+|2\rangle\langle 2|
\end{equation}

\begin{equation}\label{PP}
    P = \sum\limits_k {\sum\limits_{n = 1}^2 {\left| {{k_n}}
     \right\rangle \left\langle {{k_n}} \right|} }
       = \frac{L}{{2\pi }}\int {dk\sum\limits_{n = 1}^2
        {\left| {{k_n}} \right\rangle \left\langle {{k_n}}
         \right|} }
\end{equation}

where L is the length of waveguide, and the orthogonality
condition for P subspace vectors is

\begin{equation}\label{Ort}
    \left\langle k_n \right.\left| {k'_m} \right\rangle
     = \frac{{2\pi }}{L}\delta_{n,m}\delta \left( {k_n - k'_m} \right)
\end{equation}
where $n,m=1,2$.

The application of the method requires the continuum state vectors
to be the eigenfunctions of  Hamiltonian $H_{PP}$. This is not the
case for (\ref{P}) since $H_{PP}$ couples two vectors
$|k_1\rangle$ and $|k_2\rangle$. It is not difficult to show that
the state vectors $\left| \varphi_{i,k}  \right\rangle$ defined in
(\ref{fi}), are the eigenfunctions of $H_{PP}$ with the energies

\begin{equation}\label{En}
{E_i}/\hbar  =  - \frac{1}{2}{\omega _c} + {\omega _c}(N - 1) +
{\omega} - \frac{1}{2}{\left( { - 1} \right)^i}{\Omega _R^{(N-1)}}
\end{equation}
where $\omega$ is the frequency of incident photon.

The matrix elements of $H_{eff}$ in the $Q$ subspace is as follows

\begin{subequations}
\begin{equation}\label{Heff_11}
\left\langle {1|{H_{eff}}|1} \right\rangle  = {\omega _C}N -
\frac{1}{2}\Omega  - jN\Gamma
\end{equation}

\begin{equation}\label{Heff_22}
\left\langle {2|{H_{eff}}|2} \right\rangle  = {\omega _C}(N - 1) +
\frac{1}{2}\Omega  - j(N - 1)\Gamma
\end{equation}

\begin{equation}\label{Heff_12}
\left\langle {1|{H_{eff}}|2} \right\rangle  = \left\langle
{2|{H_{eff}}|1} \right\rangle  = \lambda \sqrt N
\end{equation}
\end{subequations}

where we introduce the width of the cavity decay rate $\Gamma =
L{\xi ^2}/v_g$. The details of the calculation of Eqs.
\ref{Heff_11}, \ref{Heff_22}, \ref{Heff_12} are given in the
Appendix \ref{A2}.

Due to the interaction of the $Q$ states (\ref{Q}) with continuum
states (\ref{P}) the former acquire the resonances whose energies
and widths become dependent on the coupling parameter $\xi$ in
Hamiltonian (\ref{H_ph}), which defines the width of the cavity
decay rate $\Gamma$. These resonances are given by the complex
roots of Eq.\ref{2qb2}. For $H_{eff}$ given by the matrix elements
(\ref{Heff_11}), (\ref{Heff_22}), (\ref{Heff_12}) this equation
reads:

\begin{multline}\label{roots}
D(z)=\left( {z/\hbar  + \frac{1}{2}\Omega - {\omega _c}N +
jN\Gamma } \right)\\\times\left( {z/\hbar  - \frac{1}{2}\Omega -
{\omega _c}\left( {N - 1} \right) + j\left( {N - 1} \right)\Gamma
} \right) - {\lambda ^2}N = 0
\end{multline}

where the complex energy $z$ is given by (\ref{En}) where the
frequency of incident photon $\omega$ is replaced by the complex
value $\widetilde{\omega}$.

Every of two $Q$ states (\ref{Q}) may decay in two ways: either to
the state $|\varphi_{1,k}\rangle$ with the energy $E_1$ or to the
state $|\varphi_{2,k}\rangle$ with the energy $E_2$.

 Accordingly, in both cases $(i=1,2)$ we obtain:

 \begin{equation}\label{DE}
    D(E_i)=(\omega-\widetilde{\omega}_{i+})(\omega-\widetilde{\omega}_{i-})
\end{equation}

where $\omega_{i\pm}$ are complex roots of the equation
(\ref{roots}).

\begin{subequations}
\begin{eqnarray}\label{roots1}
{\tilde \omega _{1, \pm }} = {\omega _C} - \frac{1}{2}\left[
{{\Omega _R^{(N-1)}} + j(2N - 1)\Gamma } \right]\nonumber\\ \pm
\frac{1}{2}\sqrt {{{\left( {\Delta  - j\Gamma } \right)}^2} +
4{\lambda ^2}N}
\end{eqnarray}

\begin{eqnarray}\label{roots2}
{\tilde \omega _{2, \pm }} = {\omega _C} + \frac{1}{2}\left[
{{\Omega _R^{(N-1)}} - j(2N - 1)\Gamma } \right]\nonumber\\ \pm
\frac{1}{2}\sqrt {{{\left( {\Delta  - j\Gamma } \right)}^2} +
4{\lambda ^2}N}
\end{eqnarray}
\end{subequations}

\begin{figure}
   \includegraphics[width=7 cm]{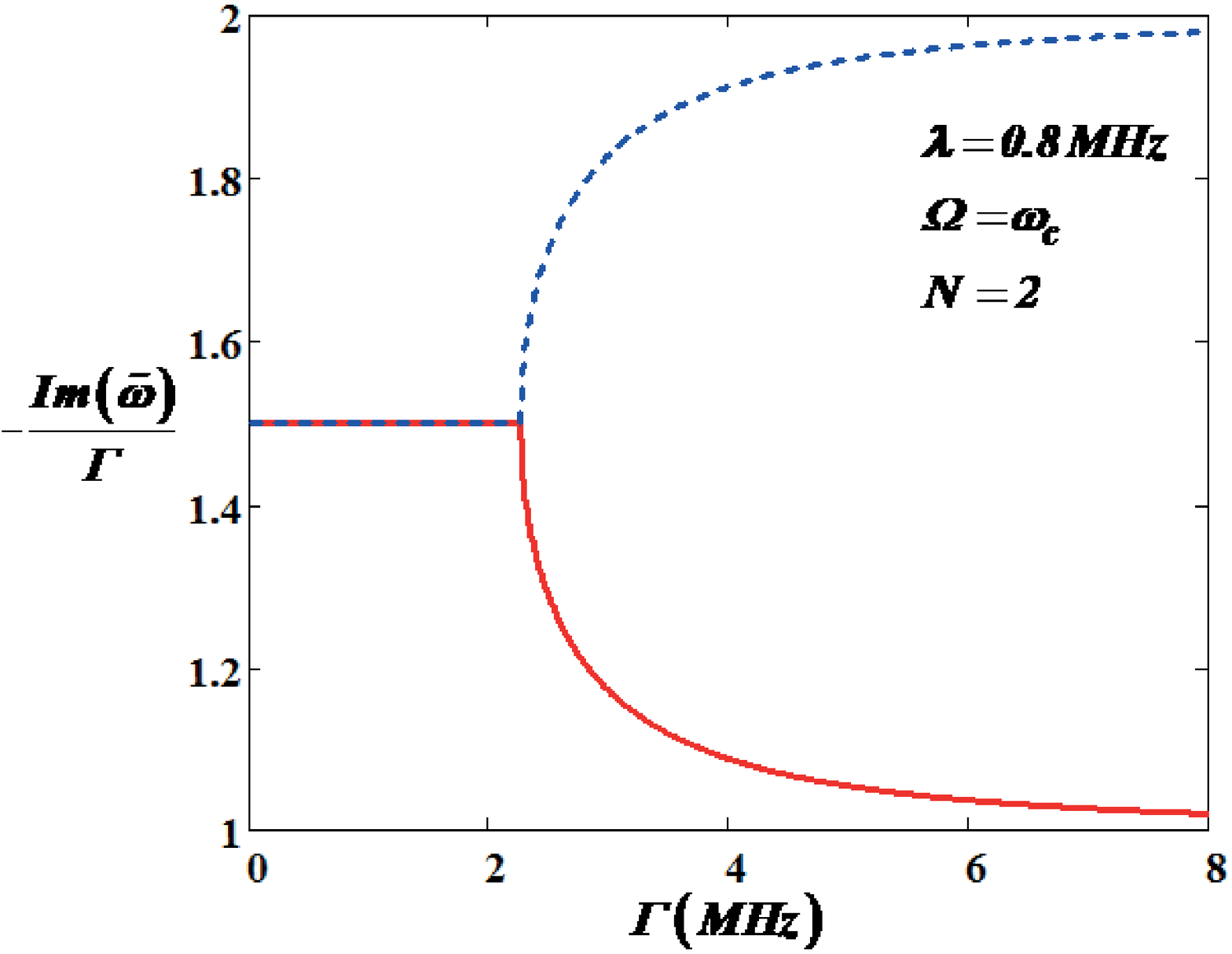}\vspace{0.5 cm}
\includegraphics[width=7 cm]{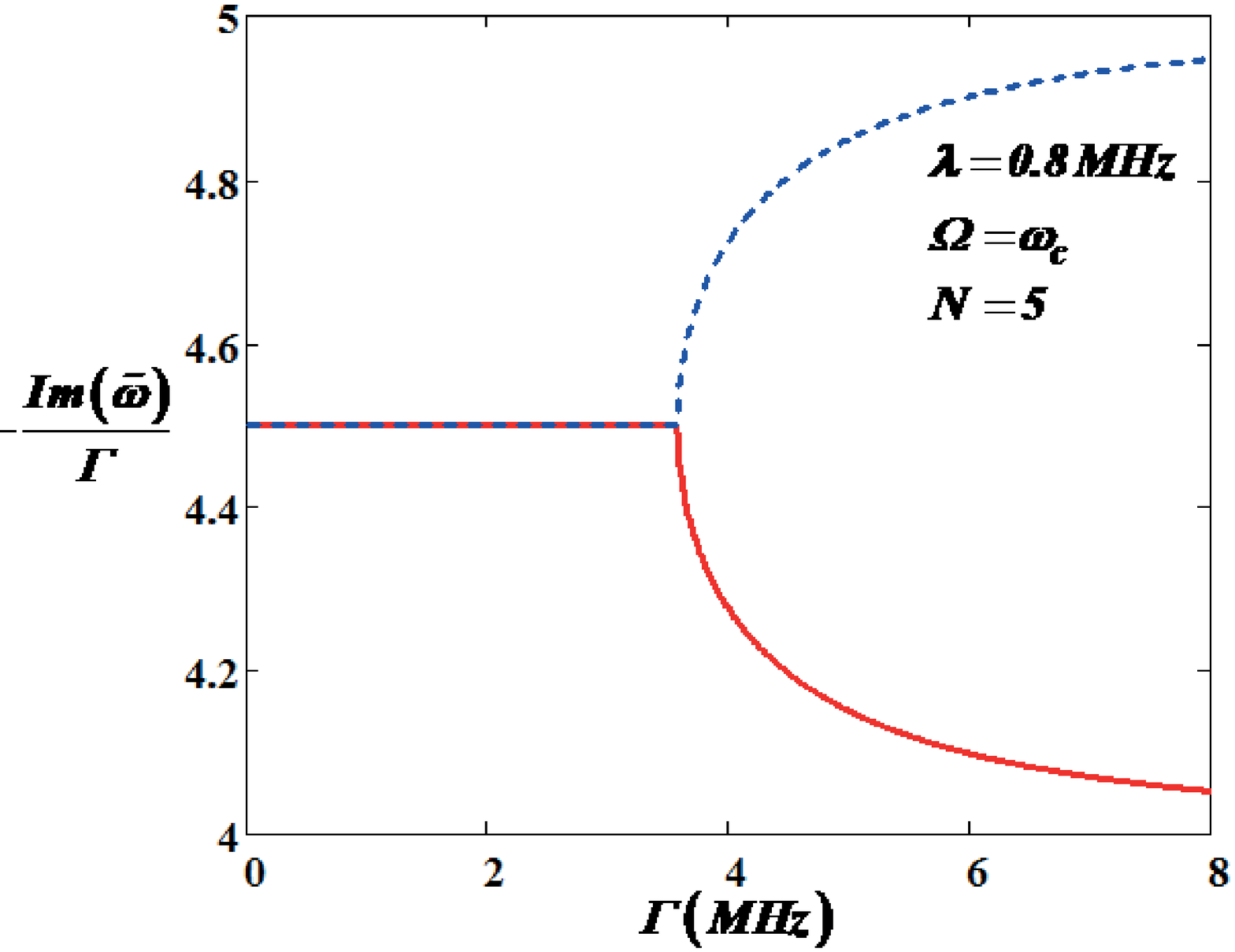}\\
  \caption{(Color online) The dependence of the resonance widths
  on the cavity decay rate $\Gamma$ for $\Delta=0$.
  For $\Gamma<2\lambda\sqrt{N}$ all widths are the same.
  The splitting starts at the point $\Gamma=2\lambda\sqrt{N}$.
  Dashed (blue) line corresponds to
  ${\tilde \omega _{1,-},\tilde \omega _{2,-}}$. Solid (red) line corresponds to
  ${\tilde \omega _{1,+},\tilde \omega _{2,+÷}}$}\label{fig3}
\end{figure}

Since ${\tilde \omega _{2, \pm }}={\tilde \omega _{1, \pm
}}+\Omega_R^{(N-1)}$, the dependence of real and imaginary part of
these resonances on $\Gamma$ is the same for both cases. The
dependence of the resonance widths on $\Gamma$ is shown in
Fig.\ref{fig3} for $\Delta=0$. The position of splitting
corresponds to the point $2\lambda\sqrt{N}=\Gamma$.

\begin{figure}
\includegraphics[width=\columnwidth,height=11 cm,keepaspectratio]{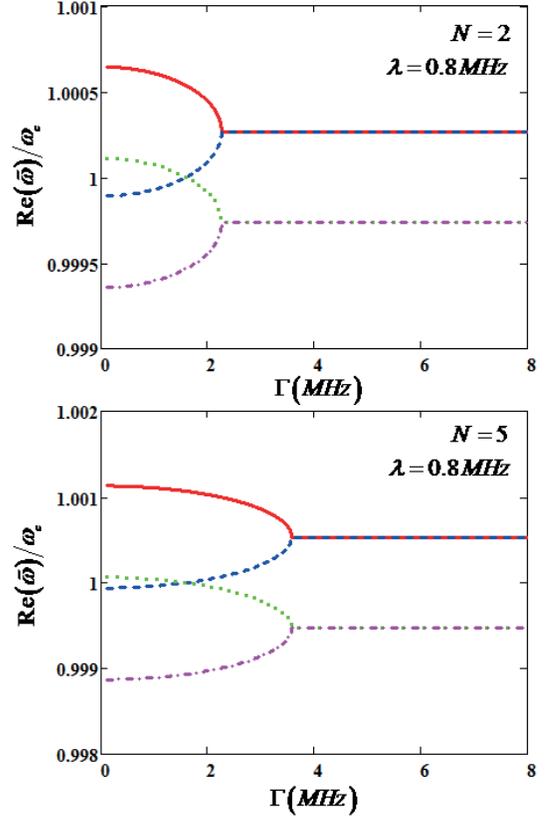}\\
\caption{(Color online) The dependence of resonance energy on
$\Gamma$ for $\Delta=0$. Upper curve corresponds to ${\tilde
\omega _{2, \pm }}$, the lower one- to ${\tilde \omega _{1, \pm
}}$. For $\Gamma>2\lambda\sqrt{N}$ the resonance energies do not
depend on $\Gamma$ and shifted by $\Omega_R$. For
$\Gamma<2\lambda\sqrt{N}$ there exist all four resonances
separately.}
  \label{fig4}
\end{figure}

The real parts of (\ref{roots1}), (\ref{roots2}) correspond to the
energy spacing between the levels of two manifolds shown in
Fig.\ref{tr11}. The transitions $\varphi_1\rightarrow\chi_2$,
$\varphi_1\rightarrow\chi_1$, $\varphi_2\rightarrow\chi_2$,
$\varphi_2\rightarrow\chi_1$ correspond to
$Re(\widetilde{\omega}_{1-})$, $Re(\widetilde{\omega}_{1+})$,
$Re(\widetilde{\omega}_{2-})$, $Re(\widetilde{\omega}_{2+})$,
respectively.

\begin{figure}
  \includegraphics[width=\columnwidth,height=11 cm,keepaspectratio]{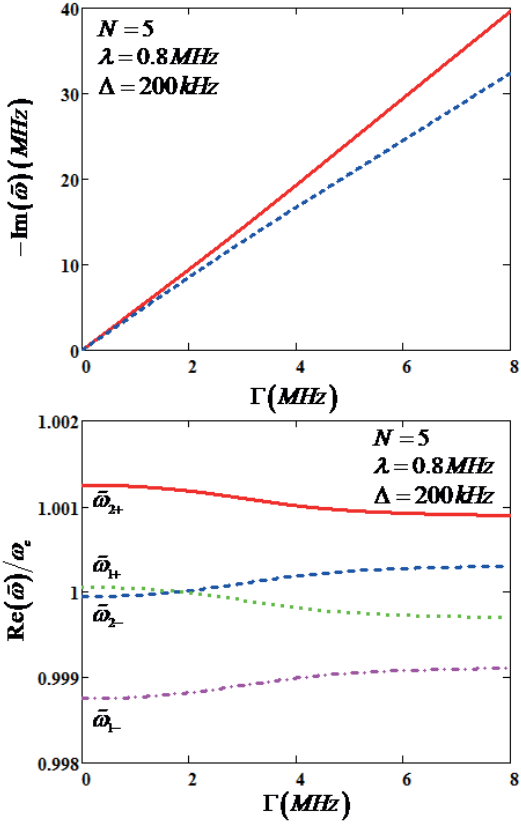}\\
  \caption{(Color online) The dependence of imaginary (upper plot) and  real (lower plot)
  parts of resonances
  on $\Gamma$ for nonzero detuning. The solid (red) curve at upper plot corresponds
  to $\widetilde{\omega}_{1,2+}$, while the dashed (blue) curve corresponds to
  $\widetilde{\omega}_{1,2-}$}.\label{Fig5}
\end{figure}

\begin{figure}
  \includegraphics[width=7 cm]{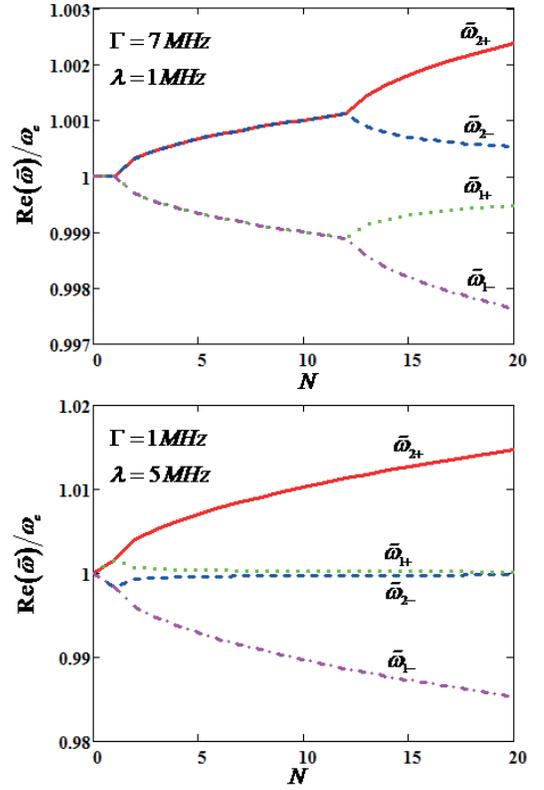}\\
  \caption{(Color online) The dependence of real parts of
  $\widetilde{\omega}_{1,\pm}$ and
  $\widetilde{\omega}_{2,\pm}$ for weak (upper plot) and strong
  (lower plot) coupling on the photon number $N$ for zero detuning. }.\label{Fig_P1}
\end{figure}

\begin{figure}
  \includegraphics[width=8cm]{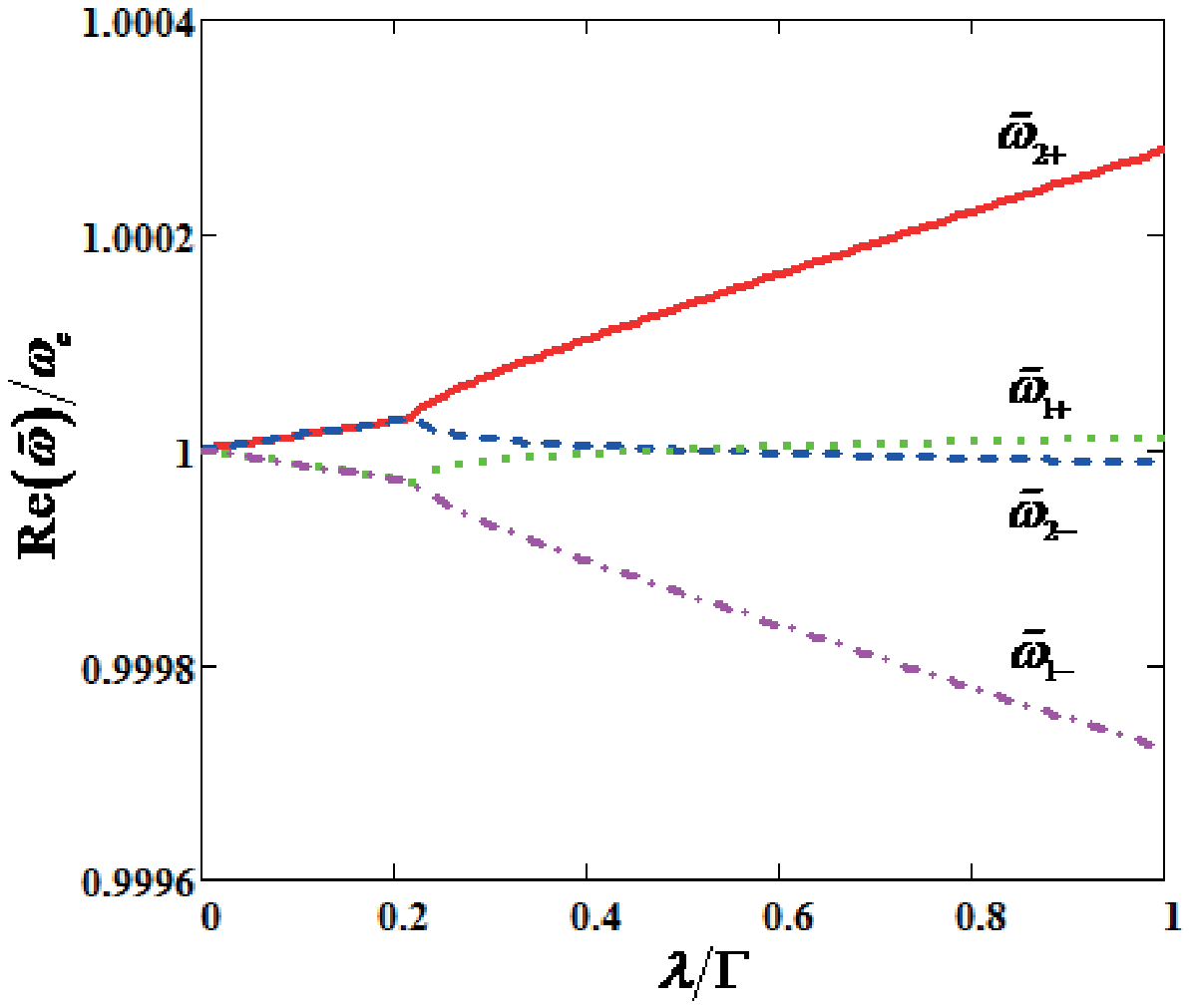}\\[0.2 cm]
  \caption{(Color online) The dependence of the positions of the resonance
  peaks on the coupling strength between qubit and
cavity photons for $\Delta=0$ and $N=5$.}\label{Re_en}
\end{figure}

Figure \ref{fig4} shows the dependence of resonance energies on
$\Gamma$ for $\Delta=0$, where for $\Gamma>2\lambda\sqrt{N}$ the
resonance energies do not depend on $\Gamma$ and shifted by
$\Omega_R^{(N-1)}$. For $\Gamma<2\lambda\sqrt{N}$ there exist all
four resonances separately. For nonzero detuning $\Delta$ the
widths are split for any $\Gamma$ as shown in the upper plot of
Fig.\ref{Fig5}. The real parts of resonance energies displays all
four components as shown in the lower plot of Fig.\ref{Fig5}. The
dependence of resonances on the photon number $N$ for weak and
strong coupling is shown in Fig.\ref{Fig_P1} for zero frequency
detuning $\Delta=0$. From (\ref{roots1}), (\ref{roots2}) we can
analyze the dependence of resonance frequencies on the coupling
strength $\lambda$.  For relatively small coupling
$\lambda/\Gamma<1/2\sqrt{N}$,
$Re(\widetilde{\omega}_{1+})=Re(\widetilde{\omega}_{1-})$ and
$Re(\widetilde{\omega}_{1+})=Re(\widetilde{\omega}_{1-})$. The
splitting begins at the point $\lambda/\Gamma=1/2\sqrt{N}$. As the
ratio $\lambda/\Gamma$ is further increased, the frequencies
(\ref{roots1}), (\ref{roots2}) scale as follows:
$Re(\widetilde{\omega}_{1-})\approx\omega_c-2\lambda\sqrt{N}$,
$Re(\widetilde{\omega}_{2+})\approx\omega_c+2\lambda\sqrt{N}$,
$Re(\widetilde{\omega}_{1+})\approx\omega_c+\lambda/2\sqrt{N},
Re(\widetilde{\omega}_{2-})\approx\omega_c-\lambda/2\sqrt{N}$.
These features are shown in Fig. \ref{Re_en} for zero detuning and
$N=5$.

As we show in Sec.V, the transmission factors scales as $1/D(E_1)$
or $1/D(E_2)$. Therefore, the resonances of these quantities,
which are given by the roots (\ref{roots1}) and  (\ref{roots2})
reflects the intrinsic properties of the cavity-qubit system. We
will see below that transmission and reflection factors are peaked
at the energies which correspond to the real parts of
(\ref{roots1}) and (\ref{roots2}).

\section{The wave function of the scattering photon}\label{MT}

The key notion for the subsequent calculation of photon
transmission and reflection is a transmission matrix

\begin{multline}\label{TM}
\left\langle {j,k'} \right|T\left| {k,i} \right\rangle \\ =
\sum\limits_{n,m = 1}^2 {\left\langle {{\varphi _{j,k'}}}
\right|{H_{PQ}}\left| n \right\rangle {R_{n,m}(E_i)}\left\langle m
\right|{H_{QP}}\left| {{\varphi _{i,k}}} \right\rangle }
\end{multline}

where the matrix $R_{m,n}(E)=(\langle
m|(E-H_{eff})|n\rangle)^{-1}$ is calculated in Appendix \ref{R}.

In our case the transmission matrix (\ref{TM}) does not depend on
the final momentum $k'$ (detail are given in Appendix). The
dependence of (\ref{TM}) on initial momentum $k$ is hidden in the
energies $E_i$ (\ref{En}), which depend on the frequency $\omega$
of incident photon.

The quantity (\ref{TM}) describes  the process where the incident
photon with momentum $k$ comes into interaction with a cavity that
was initially in the state $|\varphi_i\rangle$ and then escapes
with momentum $k'$ leaving the cavity in the state
$|\varphi_j\rangle$. Therefore, four different outcomes of this
scattering processes for transmitted probe signal are possible:
two of them correspond to elastic scattering and two of them
correspond to inelastic process with the momenta of outgoing
photon $k'=k\pm \Omega_R^{(N-1)}/v_g$ (see Fig.\ref{Fig2}).
According to these possibilities the initial state $|in\rangle$ in
(\ref{Ph1}) corresponds to either $|\varphi_{1,k}\rangle$ or
$|\varphi_{2,k}\rangle$.

\begin{multline}\label{Psi_1a}
|{\Psi _1}\rangle  = \left| {{\varphi _{1,k}}} \right\rangle  +
\sum\limits_{m,n} {\left| n \right\rangle }
{R_{nm}}({E_1})\left\langle m \right|{H_{QP}}\left| {{\varphi
_{1,k}}} \right\rangle \\ + \sum\limits_{q,i} {\frac{{\left|
{{\varphi _{i,q}}} \right\rangle }}{{{E_1}(k) - {E_i}(q) +
i\varepsilon }}\left\langle {i,q} \right|T\left| {1,k}
\right\rangle }
\end{multline}

\begin{multline}\label{Psi_2a}
|{\Psi _2}\rangle  = \left| {{\varphi _{2,k}}} \right\rangle  +
\sum\limits_{m,n} {\left| n \right\rangle }
{R_{nm}}({E_2})\left\langle m \right|{H_{QP}}\left| {{\varphi
_{2,k}}} \right\rangle \\ + \sum\limits_{q,i} {\frac{{\left|
{{\varphi _{i,q}}} \right\rangle }}{{{E_2}(k) - {E_i}(q) +
i\varepsilon }}\left\langle {i,q} \right|T\left| {2,k}
\right\rangle }
\end{multline}

From (\ref{Psi_1a}), (\ref{Psi_2a}) we obtain the photon
wavefunctions in the configuration space $\langle x|\Psi_1\rangle$
and $\langle x|\Psi_2\rangle$:

\begin{equation}\label{Phwf1}
\left\langle {x|{\Psi _1}} \right\rangle  = {e^{ikx}}\left|
{{\varphi _1}} \right\rangle  - i\Gamma {e^{ik\left| x
\right|}}{t_{11}}\left| {{\varphi _1}} \right\rangle  - i\Gamma
{e^{i\left( {k + k_R} \right)\left| x \right|}}{t_{21}}\left|
{{\varphi _2}} \right\rangle
\end{equation}

\begin{equation}\label{Phwf2}
\left\langle {x|{\Psi _2}} \right\rangle  = {e^{ikx}}\left|
{{\varphi _2}} \right\rangle  - i\Gamma {e^{ik\left| x
\right|}}{t_{22}}\left| {{\varphi _2}} \right\rangle  - i\Gamma
{e^{i\left( {k - k_R} \right)\left| x \right|}}{t_{12}}\left|
{{\varphi _1}} \right\rangle
\end{equation}
where $k_R=\Omega_R^{(N-1)}/v_g$.

The quantities $t_{ij}, i,j=1,2$ are the probability amplitudes
for the spontaneous transitions between the levels of two Rabi
doublets (see Fig.\ref{tr11}). They are related to the
transmission matrix as follows: $\left\langle {j,k'}
\right|T\left| {i,k} \right\rangle  = \left\langle {{\varphi _j}}
\right|T\left| {{\varphi _i}} \right\rangle  \equiv \xi^2t_{j,i}$.
The calculations, the details of which are given in the Appendix
\ref{A3}, yield the following expressions for the probability
amplitudes:

\begin{equation}\label{t11}
\begin{array}{l}
t_{11}  = \displaystyle\frac{1}{{4{\Omega
_R^{(N-1)}}D({E_1})}}\\[0.5 cm]
\quad\quad\quad\quad\times\left[ {N({\Omega _R^{(N-1)}} + \Delta
)\left( {2\delta  + \Delta  + {\Omega _R^{(N-1)}}} \right)}
\right.\\[0.3 cm]
 \quad\quad\quad\quad+ (N - 1)({\Omega _R^{(N-1)}} - \Delta )\left( {2\delta  - \Delta  + {\Omega _R^{(N-1)}}}
  \right)\\[0.3 cm]
\quad\quad\quad\quad+4jN(N - 1){\Omega _R^{(N-1)}}\Gamma \left. {
+ 8{\lambda ^2}N(N - 1)} \right]
\end{array}
\end{equation}

\begin{equation}\label{t21}
t_{21} =  - \frac{{\lambda \sqrt {N - 1} }}{{2{\Omega
_R^{(N-1)}}D({E_1})}}\left( {2\delta  + {\Omega _R^{(N-1)}} -
\Delta } \right)
\end{equation}

\begin{equation}\label{t22}
\begin{array}{l}
t_{22} =
 \displaystyle\frac{1}{{4{\Omega _R^{(N-1)}}D({E_2})}}\\[0.5 cm]
 \quad\quad\quad\quad\times
 \left[ {N({\Omega _R^{(N-1)}} - \Delta )
 \left( {2\delta  + \Delta  - {\Omega _R^{(N-1)}}} \right)} \right.\\[0.2 cm]
 \quad\quad\quad\quad+ (N - 1)({\Omega _R^{(N-1)}} + \Delta )
 \left( {2\delta  - {\Omega _R^{(N-1)}} - \Delta } \right)\\[0.2 cm]
\quad\quad\quad\quad\left. { + 4jN(N - 1){\Omega _R^{(N-1)}}\Gamma
- 8{\lambda ^2}N(N - 1)} \right]
\end{array}
\end{equation}

\begin{equation}\label{t12}
t_{12} =  - \frac{{\lambda \sqrt {N - 1} }}{{2{\Omega
_R^{(N-1)}}D({E_2})}}\left( {2\delta  - {\Omega _R^{(N-1)}} -
\Delta } \right)
\end{equation}

where $\delta=\omega-\omega_c$, $\Delta=\omega_c-\Omega$.

The positions of resonances are given by the points where the real
parts of the complex roots of $D(E_1)$ and $D(E_2)$ are equal to
zero. As it follows from (\ref{DE}) every quantity $t_{11},
t_{21}, t_{22}, t_{12}$ has two resonant points, while the
resonances of $t_{11}$ and $t_{21}$ (or for $t_{22}$ and $t_{12}$)
lie at the same points. It is not difficult to find this resonance
points for strong coupling ($\lambda\gg\Gamma$) and zero detuning
($\Delta=0$). The result is as follows:

\begin{subequations}
\begin{equation}\label{rest11}
    \omega_1=\omega_c+\lambda(\sqrt{N}-\sqrt{N-1})
\end{equation}
\begin{equation}\label{rest21}
    \omega_2=\omega_c-\lambda(\sqrt{N}+\sqrt{N-1})
\end{equation}
\end{subequations}

for $t_{11}$ and $t_{21}$, and

\begin{subequations}
\begin{equation}\label{rest22}
    \omega_1=\omega_c-\lambda(\sqrt{N}-\sqrt{N-1})
\end{equation}
\begin{equation}\label{rest12}
    \omega_2=\omega_c+\lambda(\sqrt{N}+\sqrt{N-1})
\end{equation}
\end{subequations}
for $t_{22}$ and $t_{12}$.

The equations (\ref{Phwf1}) and (\ref{Phwf2}) are the main results
of our paper. They have a clear physical sense. The transmission
signal (at $x>0$) consists of four waves: two elastic scattering
waves with transmission factors $T_{11}=1-i\Gamma t_{11}$,
$T_{22}=1-i\Gamma t_{22}$, and two inelastic scattering waves with
transmission factors $T_{12}=-i\Gamma t_{12}$, $T_{21}=-i\Gamma
t_{21}$. Accordingly, for reflection waves (at $x<0$) we have
$R_{ij}=-i\Gamma t_{ij}$.

For every initial state the system was before the scattering
 there are two ways for incoming photon to be scattered (see Fig.\ref{Fig2}).
 This is seen in Eqs. (\ref{Phwf1}) and
(\ref{Phwf2}) where the every scattering route is a superposition
of two final states $|\varphi_1\rangle$ and $|\varphi_2\rangle$.
The probability amplitudes $t_{11}$ (\ref{t11}), $t_{21}$
(\ref{t21}) correspond to the channels $A$ and $C$ in
Fig.\ref{Fig2}, and the amplitudes $t_{22}$ (\ref{t22}), $t_{12}$
(\ref{t12}) correspond to the channels $B$ and $D$, respectively.

It is worth noting here that the probability amplitudes in
(\ref{Phwf1}), (\ref{Phwf2}) describe different output photons.
The amplitudes $t_{11}$ and $t_{22}$ are the probabilities to find
the output photon with the same frequency as the frequency of the
input photon, while the amplitudes $t_{21}$ and $t_{12}$ are the
probabilities to find the output photon with the frequency which
is shifted from the frequency of the input photon by a Rabi
frequency $\Omega_R^{(N-1)}$.

We can show by the direct calculation that there exists an exact
condition:

\begin{equation}\label{Norm}
    |T_{ii}|^2+|1-T_{ii}|^2+2|T_{ji}|^2=1
\end{equation}
where $i,j=1,2$ and $i\neq j$ in third term in l.h.s. of
(\ref{Norm}). The left hand side of (\ref{Norm}) is a sum of
transmitted and reflected waves for every route shown in
(\ref{Phwf1}) and (\ref{Phwf2}). It is tempting to consider the
equation (\ref{Norm}) as a condition of the energy flux
conservation. However, in our case, as is seen from (\ref{Phwf1})
and (\ref{Phwf2}), the energies of the input and output photons
may be different. The condition (\ref{Norm}) reflects the
conservation of probability: after the scattering the system must
be definitely in one of the states, $|\varphi_1\rangle$ or
$|\varphi_2\rangle$.

Since for every route (\ref{Phwf1}) or (\ref{Phwf2}) there are two
outgoing photons with different frequencies we can measure
separately all transmission $T_{ij}$ (or reflection $R_{ij}$)
amplitudes.

\section{Transmission spectra}\label{TS}

As is well known the classical Mollow fluorescent spectrum
consists of three lines. However, if the number of cavity photons
is small the distance between the Rabi levels in neighbor Rabi
doublets is not equal to each other:
$\Omega_R^{(N)}>\Omega_R^{(N-1)}$. In this case the fluorescent
spectrum for two adjacent doublets will consist of four spectral
lines. These lines correspond to the spontaneous transitions
between states (see Fig.\ref{tr11}).
$|\chi_2\rangle\rightarrow\varphi_1\rangle$,
$|\chi_2\rangle\rightarrow\varphi_2\rangle$,
$|\chi_1\rangle\rightarrow\varphi_1\rangle$,
$|\chi_1\rangle\rightarrow\varphi_2\rangle$ with the corresponding
frequencies of emitting photons:
$\omega_c-\frac{1}{2}(\Omega_R^{N}+\Omega_R^{N-1})$,
$\omega_c-\frac{1}{2}(\Omega_R^{N}-\Omega_R^{N-1})$,
$\omega_c+\frac{1}{2}(\Omega_R^{N}-\Omega_R^{N-1})$,
$\omega_c+\frac{1}{2}(\Omega_R^{N}+\Omega_R^{N-1})$.

The result of our study shows that we obtain the same frequencies
for transmitted photons when studying the scattering of a single
photon in 1D geometry via the system shown in Fig.\ref{SC}. In
addition, we obtained the probability amplitudes (expressions
(\ref{t11}), (\ref{t21}), (\ref{t22}), (\ref{t12})) for
spontaneous transitions between levels of two Rabi doublets (see
Fig.\ref{tr11}). Below we illustrate the application of our
results to the transmission spectra for $N=2$ for strong resonance
coupling when the distance between Rabi levels within $N$ manifold
are given by $\Omega_R^{(N)}$ (\ref{rabi}).

Having in mind to study the effects of adding to a cavity one
extra photon we find the transmission and reflection factors for
$N=1$ where we have either one photon in a waveguide and no photon
in a cavity with a qubit being in its ground state or no photons
in a waveguide and one photon in a cavity. In this case, as is
seen from (\ref{t11})-(\ref{t22}) the only quantity which is
different from zero is $t_{11}$, so that for transmission and
reflection we obtain:

\begin{equation}\label{TN1}
T_{11}^{(N = 1)} = \frac{{\left( {\omega  - {\omega _ + }}
\right)\left( {\omega  - {\omega _ - }} \right)}}{{\left( {\omega
- {\omega _ + }} \right)\left( {\omega  - {\omega _ - }} \right) +
j\Gamma \left( {\omega  - \Omega } \right)}}
\end{equation}

\begin{equation}\label{RN1}
R_{11}^{(N = 1)} = \frac{{ - j\Gamma \left( {\omega  - \Omega }
\right)}}{{\left( {\omega  - {\omega _ + }} \right)\left( {\omega
- {\omega _ - }} \right) + j\Gamma \left( {\omega  - \Omega }
\right)}}
\end{equation}

where
\begin{equation}\label{fr11}
{\omega _ \pm } = \frac{1}{2}\left( {{\omega _c} + \Omega }
\right) \pm \frac{1}{2}\Omega_R^{(1)}
\end{equation}

The expressions (\ref{TN1}) and (\ref{RN1}) coincide with those
known from the literature \cite{Shen09}. We have here two
resonances at the frequencies $\omega_\pm$ with the distance
between them being equal to Rabi frequency $\Omega_R^{(1)}$.

If we add one extra photon to the system we will also have two
resonances for every route (\ref{Phwf1}) or (\ref{Phwf2}). But the
picture is drastically different from the $N=1$ case. For example,
if before scattering the system is in $|\varphi_1\rangle$ state,
then each of the amplitudes $t_{11}$ and $t_{21}$ in (\ref{Phwf1})
has two resonances at the same frequencies. The first resonance at
$\omega_c-\frac{1}{2}(\Omega_R^{(2)}+\Omega_R^{(1)})$ corresponds
to the transition from the state $|\varphi_1\rangle$ to the state
$|\chi_2\rangle$ which subsequently decays either to the initial
state $|\varphi_1\rangle$ (the probability  of this process is
given by the amplitude $t_{11}$ in (\ref{Phwf1})) or to the state
$|\varphi_2\rangle$ with the probability  being given by the
amplitude $t_{21}$. The second resonance at
$\omega_c+\frac{1}{2}(\Omega_R^{(2)}-\Omega_R^{(1)})$ corresponds
to the transition from the state $|\varphi_1\rangle$ to the state
$|\chi_1\rangle$ which subsequently decays either to the initial
state $|\varphi_1\rangle$ with the probability $t_{11}$ or to the
state $|\varphi_2\rangle$ with the probability $t_{21}$.
Therefore, we see that each resonance corresponds to two outgoing
photons: the frequency of the first photon is equal to the input
frequency, the frequency of the second photon is increased as
compared with the first one by the amount $\Omega_R^{(1)}$. Since
the frequencies of these two photons are different, they can be
detected separately and independently of each other.

\begin{figure}
  \includegraphics[width=8cm]{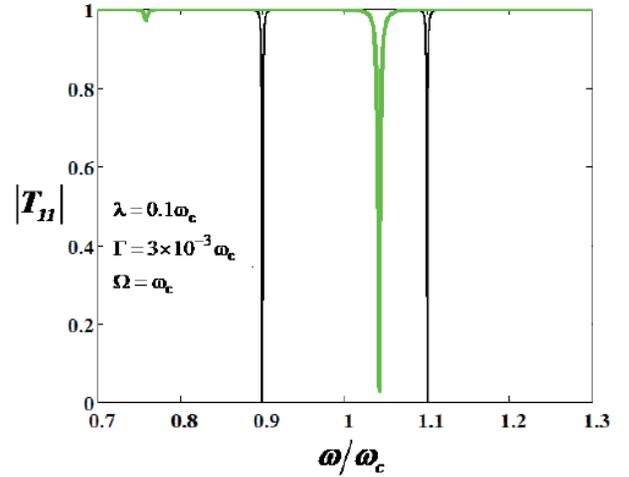}\\
  \caption{(Color online) Comparison of transmissions $T_{11}$
  for $N=1$ (black, thin line) and $N=2$ (green, thick line)
   for strong resonant coupling.}\label{N1+N2}
\end{figure}

In Fig.\ref{N1+N2} we compare the transmission coefficients
$T_{11}$ for $N=1$ and $N=2$ as a function of the frequency of
incident photon for the case of strong resonant coupling:
$\lambda\gg\Gamma$, $\omega_C=\Omega$. Two dips which are
symmetric relative to $\omega_C$ are calculated from expression
(\ref{TN1}). These dips are located at
$\omega_C\pm\Omega_R^{(1)}$. The addition of one extra photon
gives rise to the appearance of two dips, which results from the
excitation of the level $|\varphi_1\rangle$. These dips are
calculated from (\ref{t11}). A shallow dip, which is located at
the frequency
$\omega_c-\frac{1}{2}(\Omega_R^{(2)}+\Omega_R^{(1)})$ corresponds
to the transition $|\varphi_1\rangle\rightarrow
|\chi_2\rangle\rightarrow |\varphi_1\rangle$, while a deep dip,
which is located at the frequency
$\omega_c+\frac{1}{2}(\Omega_R^{(2)}-\Omega_R^{(1)})$ corresponds
to the transition $|\varphi_1\rangle\rightarrow
|\chi_1\rangle\rightarrow |\varphi_1\rangle$. The distance between
two dips is equal to $\Omega_R^{(2)}$. For both cases the
frequency of outgoing photons is equal to the frequency of the
input photon.

\begin{figure}
  \includegraphics[width=8 cm]{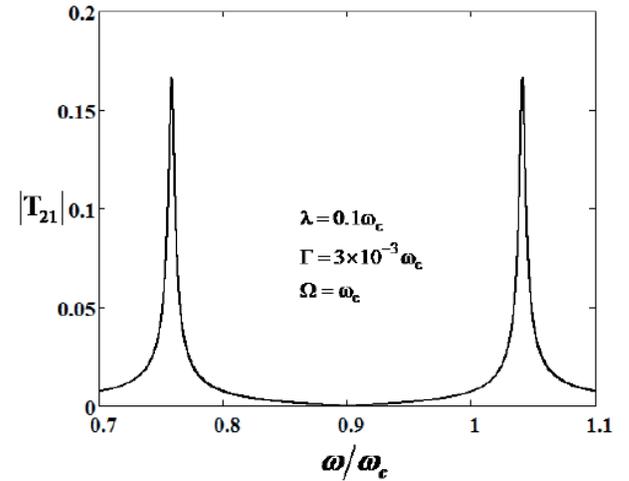}\\
  \caption{Inelastic transmission spectrum for $N=2$ and strong resonant coupling
  after the excitation of
  $|\varphi_1\rangle$ state. The outgoing photon leaves the cavity with the
  increased frequency $\omega+\Omega_R^{(1)}$.}\label{Transm21}
\end{figure}

In Fig.\ref{Transm21} we show the transmission spectrum which is
given by the amplitude $t_{21}$ in (\ref{Phwf1}). Here the
resonance points are the same as those in Fig.\ref{N1+N2},
however, the outgoing photon has the
  increased frequency $\omega+\Omega_R^{(1)}$. After the
  scattering the cavity is being left in the state
  $|\varphi_2\rangle$. The left peak in Fig.\ref{Transm21}
  corresponds to transitions $|\varphi_1\rangle\rightarrow
|\chi_2\rangle\rightarrow |\varphi_2\rangle$ with the frequency of
outgoing photon
$\omega=\omega_c-\frac{1}{2}(\Omega_R^{(2)}-\Omega_R^{(1)})$. The
right peak corresponds to transitions
$|\varphi_1\rangle\rightarrow |\chi_1\rangle\rightarrow
|\varphi_2\rangle$ with the frequency of outgoing photon
$\omega=\omega_c+\frac{1}{2}(\Omega_R^{(2)}+\Omega_R^{(1)})$.

\begin{figure}
  \includegraphics[width=8 cm,height= 5 cm]{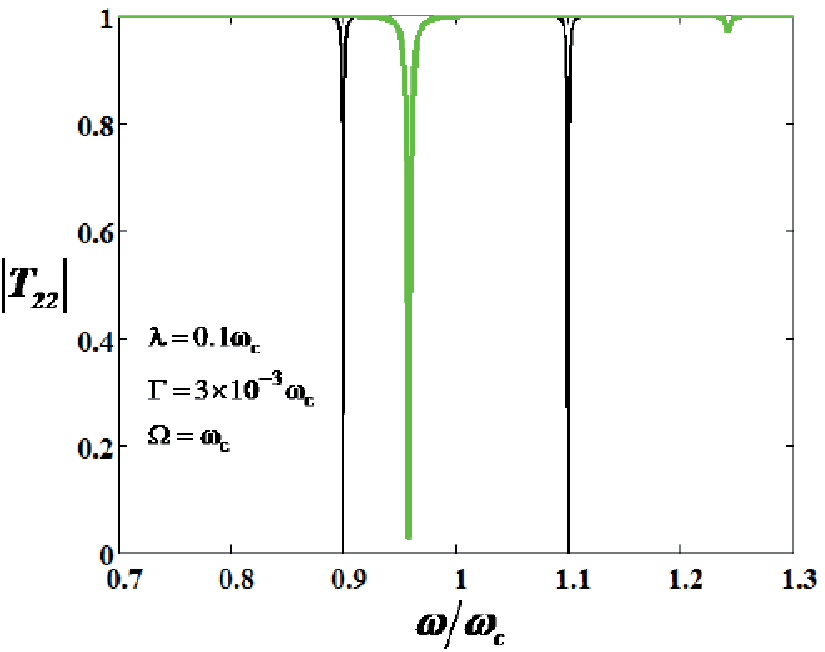}\\[0 cm]
  \caption{(Color online) Comparison of transmissions $T_{11}$
  for $N=1$ (black, thin line) and $T_{22}$ for $N=2$ (green, thick line)
   for strong resonant coupling.}\label{Transm22}
\end{figure}

If initially the system is in the state $|\varphi_2\rangle$, the
scattering wave function is given by (\ref{Phwf2}). The resonance
points are being shifted on the frequency axis to the right by
$\Omega_R^{(1)}$. The first resonance at
$\omega_c-\frac{1}{2}(\Omega_R^{(2)}-\Omega_R^{(1)})$ corresponds
to the transition $|\varphi_2\rangle\rightarrow |\chi_2\rangle$
while the second one at
$\omega_c+\frac{1}{2}(\Omega_R^{(2)}+\Omega_R^{(1)})$ corresponds
to the transition $|\varphi_2\rangle\rightarrow |\chi_1\rangle$.
Each of these excitations then decays either to the initial state
$|\varphi_2\rangle$ with the probability amplitude $t_{22}$ or to
the state $|\varphi_1\rangle$ with the probability amplitude
$t_{12}$. The transmission spectrum for $N=2$ for the case when
the system is left after scattering in the state
$|\varphi_2\rangle$ is shown in Fig.\ref{Transm22}. This picture
is similar to that shown in Fig.\ref{N1+N2}. A deep dip, which is
located at the frequency
$\omega_c-\frac{1}{2}(\Omega_R^{(2)}-\Omega_R^{(1)})$ corresponds
to the transition $|\varphi_2\rangle\rightarrow
|\chi_2\rangle\rightarrow |\varphi_2\rangle$, while a shallow dip,
which is located at the frequency
$\omega_c+\frac{1}{2}(\Omega_R^{(2)}+\Omega_R^{(1)})$ corresponds
to the transition $|\varphi_2\rangle\rightarrow
|\chi_1\rangle\rightarrow |\varphi_2\rangle$. The distance between
two dips is equal to $\Omega_R^{(2)}$. For both cases the
frequency of outgoing photons is equal to the frequency of the
input photon.

\begin{figure}
  \includegraphics[width=8 cm]{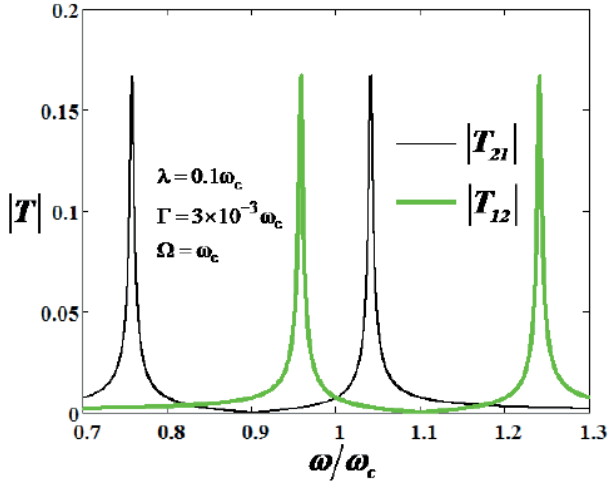}\\
  \caption{Color online. Inelastic transmission spectra for $N=2$ and strong resonant coupling
  after the excitation of
   the state $|\varphi_1\rangle$ (thin, black line) and the state $|\varphi_2\rangle$
   (thick, green line).}\label{T1221}
\end{figure}

In Fig.\ref{T1221} we show in one plot the transmission spectra
which are given by the amplitudes $t_{21}$ in (\ref{Phwf1}) and
$t_{12}$ in (\ref{Phwf2}). The black thin lines show the
transmission spectrum when the system was initially in the state
$|\varphi_1\rangle$ and after scattering was left in the state
$|\varphi_2\rangle$ with the outgoing photon with the frequency
increased by $\Omega_R^{(1)}$. These spectrum is the same as is
shown in Fig.\ref{Transm21}. The green thick lines in
Fig.\ref{T1221} show the transmission spectrum when the system was
initially in the state $|\varphi_2\rangle$ and after scattering
was left in the state $|\varphi_1\rangle$ with the outgoing photon
with the frequency reduced by $\Omega_R^{(1)}$. The left peak of
this spectrum corresponds to the excitation of the transition
$|\varphi_2\rangle\rightarrow |\chi_2\rangle$ at the frequency of
ingoing photon
$\omega_c-\frac{1}{2}(\Omega_R^{(2)}-\Omega_R^{(1)})$. The state
$|\chi_2\rangle$ then decays to the state $|\varphi_1\rangle$ with
the frequency of outgoing photon
$\omega_c-\frac{1}{2}(\Omega_R^{(2)}+\Omega_R^{(1)})$. The right
peak corresponds to the excitation of the transition
$|\varphi_2\rangle\rightarrow |\chi_1\rangle$ by the ingoing
photon with the frequency
$\omega_c+\frac{1}{2}(\Omega_R^{(2)}+\Omega_R^{(1)})$. The state
$|\chi_1\rangle$ then decays to the state $|\varphi_1\rangle$ with
the frequency of outgoing photon
$\omega_c+\frac{1}{2}(\Omega_R^{(2)}-\Omega_R^{(1)})$.

\subsection{Comparison with the experiment}\label{experFink}
\subsubsection{The frequencies of the probing and detected photons are the same}

\begin{figure}
  \includegraphics[width=8 cm]{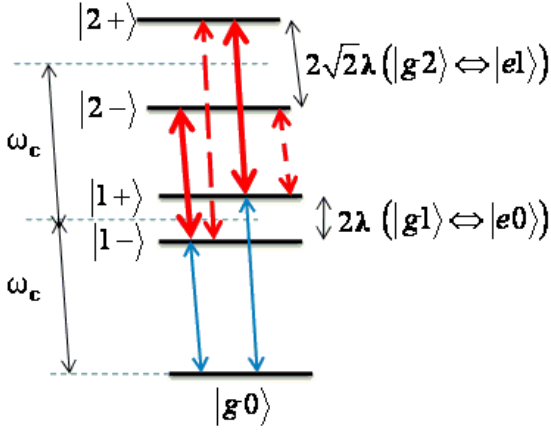}\\
  \caption{Color on line. Level diagram of a resonant cavity QED system for $N=2$ \cite{Fink08}.
Thin blue arrows between $|g0\rangle$ and $|1\pm\rangle$ levels
are responsible for the vacuum Rabi mode splitting which is shown
by black dips in Fig.\ref{N1+N2} and Fig.\ref{Transm22}. Thick
solid (red) arrows correspond to the green dips, and the solid
dashed (red) lines correspond to shallow dips in Fig.\ref{N1+N2}
and Fig.\ref{Transm22}.}\label{Fink}
\end{figure}

We show here that our results shown in Fig.\ref{N1+N2} and
Fig.\ref{Transm22} correspond to those measured in \cite{Fink08},
where atom– photon superposition states involving up to two
photons have been studied, using a spectroscopic pump and probe
technique. The experiments have been performed in a circuit QED
setup, in which a superconducting qubit of transmon type has been
embedded in a high-quality on-chip microwave cavity so that the
frequency of the input (probing) photon and that of the output
(detected) photon coincides. The level diagram of this system for
$N=2$ is shown in Fig.\ref{Fink}.

The measurements were performed on resonance ($\omega_c=\Omega$)
and under conditions of very strong coupling ($\lambda \gg
\gamma$) where $\gamma$ is the qubit dephasing rate. The first and
second Rabi doublets in Fig.\ref{Fink} are due to the
hybridization of the  bare qubit-photon states $|g1\rangle$,
$|e0\rangle$, and $|g2\rangle$, $|e1\rangle$, respectively.

Our scheme is different from that of Ref.\cite{Fink08} in that we
consider here a side coupled configuration with the open broad
band waveguide while in \cite{Fink08} the measurements have been
performed for direct coupled configuration with a high-$Q$
waveguide. However, the side coupled transmission coefficients
$T_{11}$ and $T_{22}$ can be transformed to direct couple ones by
a simple transformation \cite{Shen09}. The transmission spectra
for direct coupling is equal to the side coupled reflection
spectra: $T_{ii}^{dc}=1-T_{ii}, (i=1,2)$. Hence,
$T_{11}^{dc}=j\Gamma t_{11}$, $T_{22}^{dc}=j\Gamma t_{22}$, where
$t_{11}$, $t_{22}$ are given in (\ref{t11}), (\ref{t22}),
respectively. Therefore, the transmission spectra shown in Fig.4b
and Fig.4d in \cite{Fink08} are the mirror reflection of the
spectra shown in Fig.\ref{N1+N2} and Fig.\ref{Transm22},
respectively. Two dips in these figures which are symmetric
relative to $\omega_c$ are the signature of vacuum Rabi mode
splitting. For on resonant strong coupling these dips are located
at $\omega=\omega_c\pm\lambda$ and correspond to the transitions
between ground state $|g0\rangle$ and the states $|1+\rangle$ and
$|1-\rangle$ (blue thin lines in Fig.\ref{Fink}). For on resonance
strong coupling conditions these dips give a full extinction of
transmitted signal. However, if the bandwidth of the uncoupled
waveguide is much smaller than the Rabi mode splitting the
extinction can be very small (Fig.4b in \cite{Fink08}).

The original idea in \cite{Fink08} was to measure the splitting of
second Rabi doublet. By populating the levels $|1+\rangle$ or
$|1-\rangle$ with a single photon they probed the transitions
between $|1\pm\rangle$ and $|2\pm\rangle$ levels. The transitions
$|1+\rangle\rightarrow|2+\rangle$, and
$|1+\rangle\rightarrow|2-\rangle$, are described by the
transmission amplitudes $T_{11}$, while the transitions
$|1-\rangle\rightarrow|2-\rangle$ and,
$|1-\rangle\rightarrow|2+\rangle$ are described by the
transmission amplitudes $T_{22}$. The deep dips which are shown by
green lines in Fig.\ref{N1+N2} and Fig.\ref{Transm22} lie between
vacuum Rabi modes lines. These dips which are located at the
frequencies $\omega=\omega_c+(\sqrt2-1)\lambda$,
$\omega=\omega_c-(\sqrt2-1)\lambda$ and correspond to the
transitions $|1+\rangle\rightarrow|2+\rangle$ and
$|1-\rangle\rightarrow|2-\rangle$ were observed in
\cite{Fink08}(Fig.4b,d). However, they failed to observe the
transitions $|1+\rangle\rightarrow|2-\rangle$ and
$|1-\rangle\rightarrow|2+\rangle$ which are shown by dashed red
lines in Fig.\ref{Fink}. As was noted in \cite{Fink08}, the
amplitudes of these transitions were very small to be observed.
These amplitudes can be seen as shallow dips in Fig.\ref{N1+N2}
and Fig.\ref{Transm22}. Using the data from \cite{Fink08}:
$\omega_c/2\pi=6.94$ MHz, $\lambda/2\pi=154$ MHz,
$\Gamma/2\pi=0.9$ MHz, we find from Eqs. \ref{t11}, \ref{t22} the
ratio of the amplitudes of the shallow dip to that of the main
dip. For both cases shown in Fig.\ref{N1+N2} and
Fig.\ref{Transm22} this ratio is approximately equal to $3\times
10^{-3}$.


\subsubsection{The frequencies of the probing and detected photons are different}

It is important that in \cite{Fink08} the frequencies of the input
and output photons were the same. Thus, as we show above, the
experimental results in \cite{Fink08} can be explained by the
amplitudes $t_{11}$ and $t_{22}$ in Eqs. \ref{Phwf1}, \ref{Phwf2}.

However,  the Eqs. \ref{Phwf1}, \ref{Phwf2} predict another effect
which at the best of our knowledge has not been observed in single
photon experiments. We mean the registration of the output photon
with a frequency shifted from that of the input photon by a Rabi
frequency $\Omega_R^{(N-1)}$. The amplitudes responsible for this
process are given by the quantities $t_{21}$ (\ref{t21}) and
$t_{12}$ (\ref{t12}). The corresponding resonances are shown in
Fig.\ref{T1221}. The resonance frequencies in this figure
correspond to the frequencies of the input photons which excite
the transition in the cavity, but the frequency of the outgoing
photons is different. For example, two peaks in Fig.\ref{Transm21}
correspond to the excitation of transitions (see Fig.\ref{Fink})
$|1+\rangle\rightarrow |2-\rangle$ (left peak) and
$|1+\rangle\rightarrow |2+\rangle$ (right peak) with subsequent
decay to the state $|1-\rangle$ ($|2\pm\rangle\rightarrow
|1-\rangle$) leaving the output photon with the frequency
increased by $2\lambda$. Therefore, the amplitude of, for example,
the left peak in  Fig.\ref{Transm21} should be interpreted as the
probability to find the output photon with the frequency
$\omega_c-(\sqrt{2}-1)\lambda$ if the frequency of the input
photon is $\omega_c-(\sqrt{2}+1)\lambda$.

\begin{figure}
  \includegraphics[width=8 cm]{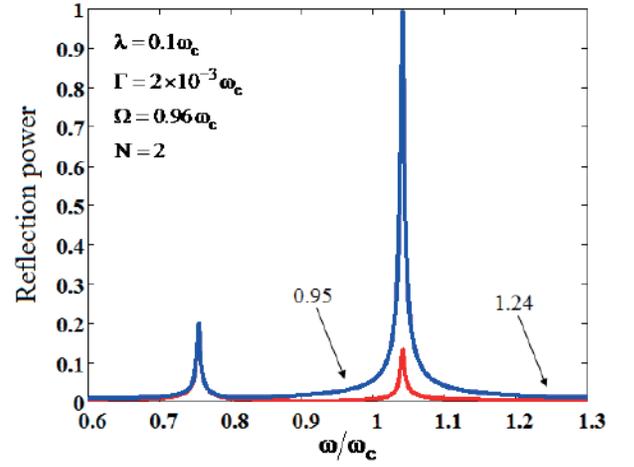}\\
  \caption{Reflection spectrum for broadband single photon detection. Black
  arrows show the frequency of the output photon the frequency of which is
  greater than the frequency of the input photon by the value of Rabi splitting
  $\Omega_R^{(1)}$. See text for details.} \label{Refl}
\end{figure}

The detection of the output photons with the frequency different
from that of the input photons can be realized using a vector
network analyzer at the output of a broadband (low $Q$) waveguide.
In order to detune from the input photons it is better to measure
the reflected spectra. For broadband waveguide the reflected
coefficient is given by the quantity

\begin{equation}\label{Rfl}
{|R_i|} \equiv \left| {\left\langle {x|{\Psi _i}} \right\rangle  -
{e^{ikx}}\left| {{\varphi _i}} \right\rangle } \right| = \Gamma
\sqrt {{{\left| {{t_{ii}}} \right|}^2} + {{\left| {{t_{ji}}}
\right|}^2}}
\end{equation}
where $i,j=1,2$ and $i\neq j$ in $t_{ji}$ in r.h.s of
(\ref{Refl}). The quantities $R_1$ and $R_2$ correspond to the
preliminary populated levels $|1+\rangle$ and $|1-\rangle$,
respectively.

The reflection spectrum for the case when the level $|1+\rangle$
is preliminary populated is shown in Fig.\ref{Refl} by a thick
blue line. The left peak at the point $\omega/\omega_c\approx
0.75$ is formed mainly by the contribution of $t_{21}$. As we
explained before, this peak gives the probability to find the
output photon at the frequency increased by $2\lambda$. This point
is shown by the left arrow in Fig.\ref{Refl}. A central large blue
peak is formed mainly by the contribution of $t_{11}$. It means
that at this input frequency $\omega\approx 1.04\omega_c$ we
observe the output photon with the same frequency. However, a
small contribution of $t_{21}$ to the central peak (shown by thin
red line peak at $\omega/\omega_c\approx 1.04$) results in the
output photon at the frequency $\omega\approx 1.24\omega_c $
(shown by the right arrow in Fig.\ref{Refl}.

The same picture exists for the case when the level $|1-\rangle$
is preliminary populated. Here the reflection is given by the
quantity $R_2$, and the output photons with the frequency
decreased by $2\lambda$ can be observed.

\section{A signature of the photon blockade in the transmission
spectra}\label{phb}

A concept of the photon blockade, in which transmission of only
one photon through a system is possible while excess photons are
absorbed or reflected, was first proposed in \cite{Imam97}. Since
then there have been published the plethora of papers devoted to
this phenomenon in different atom-cavity systems (see, for
example, recent papers \cite{Deng17, Baj13} and references there
in). The photon blockade is observed when the atom- photon
interaction results in the energy spectrum with a nonlinear
dependence on the number of cavity photons $n$. It can be either
 Kerr- type $n^2$ nonlinearity when the resonance frequency is
largely detuned from the qubit energy (so called, a dispersive
photon blockade \cite{Hoffman11}) or the resonant photon blockade
with Jaynes- Cummings $\sqrt{n}$ dependence \cite{Birnbaum05}. The
photon blockade are usually investigated using the correlation
function measurements of the photon statistics at the cavity
output \cite{Birnbaum05, Lang11}. Alternatively, the signature of
the photon blockade can be found as staircase pattern in the
dependence of transmitted power on the incident photon bandwidth
\cite{Hoffman11}.

Below we show the signature of the photon blockade in the
transmission of a single photons one-by-one through a waveguide
side coupled to the resonance cavity with a two-level atom (see
Fig.\ref{SC}). In our scheme the photon blockade manifests as the
transmission of a photon at some frequency $\omega$ if the
preceding photon with the same frequency $\omega$ have been
captured by the cavity. Or, alternatively, it may be observed at
the input:  if the input photon at some frequency $\omega$ is
captured by the cavity, we, first, observe the reflected signal,
and, second, the following photon with the same frequency passes
through waveguide producing no reflected signal.

\begin{figure}
  \includegraphics[width=8 cm]{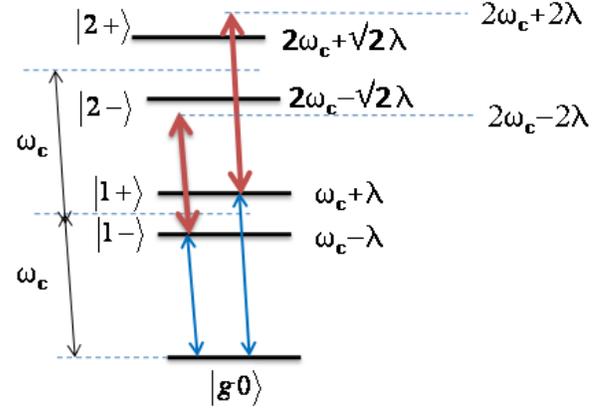}\\
  \caption{Color online.  The level structure for on resonance strong coupling limit.
  Photon blockade manifests as the suppression of two- photon absorption for a probe field of frequency
  $\omega_p=\omega_c-\lambda$  or $\omega_p=\omega_c+\lambda$ (thick red arrows)
  tuned to excite the transition $|g0\rangle\rightarrow|1-\rangle$ or
$|g0\rangle\rightarrow|1+\rangle$. The frequency gap in both cases
is equal to $(2-\sqrt{2})\lambda$.}\label{block}
\end{figure}

Even if initially there is no photons in a cavity, the first input
photon with the frequency $\omega=\Omega$ is blocked to enter the
cavity, it is completely transmitted as it follows from
Eq.\ref{TN1}. It can be captured by the cavity with simultaneous
appearance of reflected signal only if its frequency is equal to
$\omega_{\pm}$ (see Eq.\ref{RN1}). The adding of a second photon
with the frequencies $\omega_{\pm}$ cannot excite the cavity since
there are no appropriate energy levels in the cavity with two
photons with the  energies $2\hbar\omega_{\pm}$ as it is shown in
Fig.\ref{block}. There is a frequency gap within which a second
photon cannot be captured by a cavity.

Below we show that in our scheme the photon blockade appears as
the staircase pattern in the dependence of the reflected power on
the detector bandwidth $\Delta\omega$ centered at $\omega_c$.
First, we excite the cavity by a single photon with the energy
corresponding to one of the hybridized level ($|N_p-\rangle$ or
$|N_p+\rangle$) . This level successively undergoes one- photon
decays to lower hybridized states. Every of these transitions
produces a reflected signal at the corresponding frequency. Hence,
under repeated excitation of the levels $|N_p\pm\rangle$ we obtain
a reflected power as a discrete number of peaks, which number
depends on $N_p$.

\begin{figure}
  \includegraphics[width=8 cm]{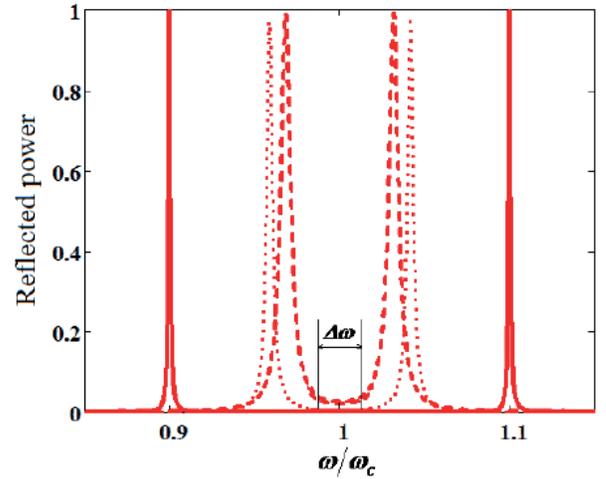}\\
  \caption{Reflected power spectral function $W_N(\omega)$ for
  $N=1$ (solid lines), $N=2$ (dotted line), and $N=3$ (dashed
  lines). The calculations is made for $\omega_c=\Omega=3$ GHz,
$\lambda=0.1\omega_c, \Gamma=2.66\times 10^{-3}\omega_c$.
$\Delta\omega$ is a variable bandwidth of the detector
.}\label{RPW}
\end{figure}

Therefore, we define the reflected power in the detector bandwidth
$\Delta\omega$ as follows:

\begin{equation}\label{RP}
{P_R}({N_p},\Delta \omega ) = \sum\limits_{N = 1}^{{N_p}}
{\int\limits_{{\omega _c} - \Delta \omega }^{{\omega _c} + \Delta
\omega } {{{\left| {W_N(\omega)} \right|}^2}d\omega } }
\end{equation}

where a spectral function $W_N(\omega)=|R_1+R_2|^2$ for $N>1$ with
$R_1, R_2$ being defined in (\ref{Rfl}), and
$W_1(\omega)=R_{11}^{(N=1)}$, where $R_{11}^{(N=1)}$ is defined in
(\ref{RN1}). For $N>1$ the quantity $W_N(\omega)$ corresponds to
the transitions $|N,\pm\rangle\rightarrow |N-1,\pm\rangle$, while
$W_1(\omega)$ describes the transitions to the ground state
$|1,\pm\rangle\rightarrow |g,0\rangle$.

The example of reflected power spectrum for first three $N'$s is
shown in Fig.\ref{RPW}. It is seen that as $N$ is increased the
width of resonance lines is also increased, which is
understandable from the inspections of expressions (\ref{roots1})
and (\ref{roots2}).

\begin{figure}
  \includegraphics[width=8 cm]{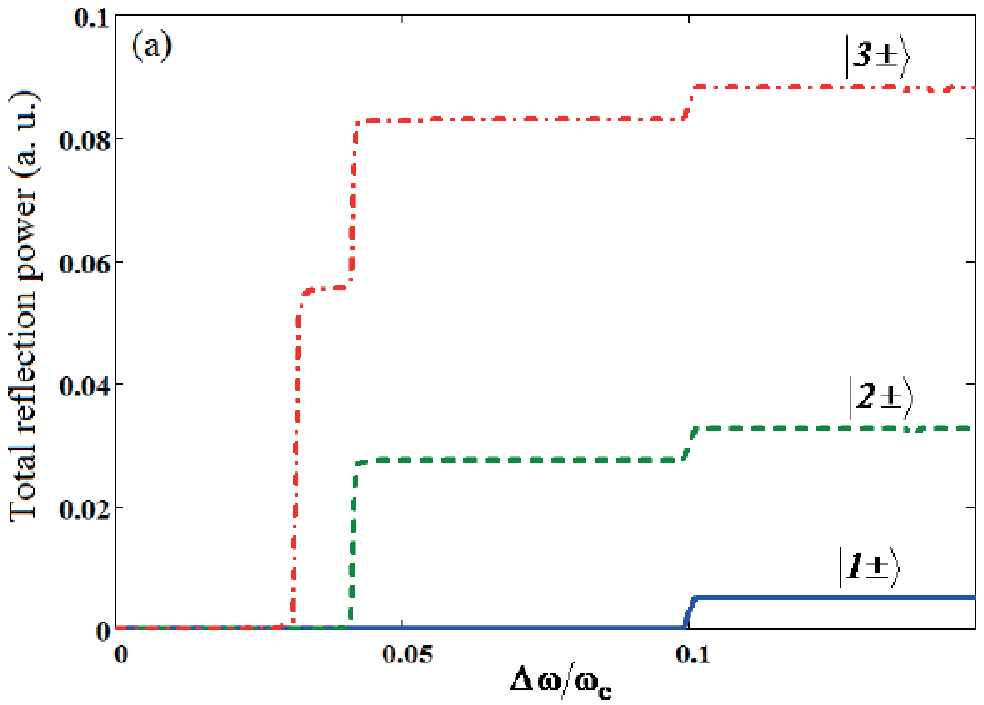}\\
\includegraphics[width=8 cm]{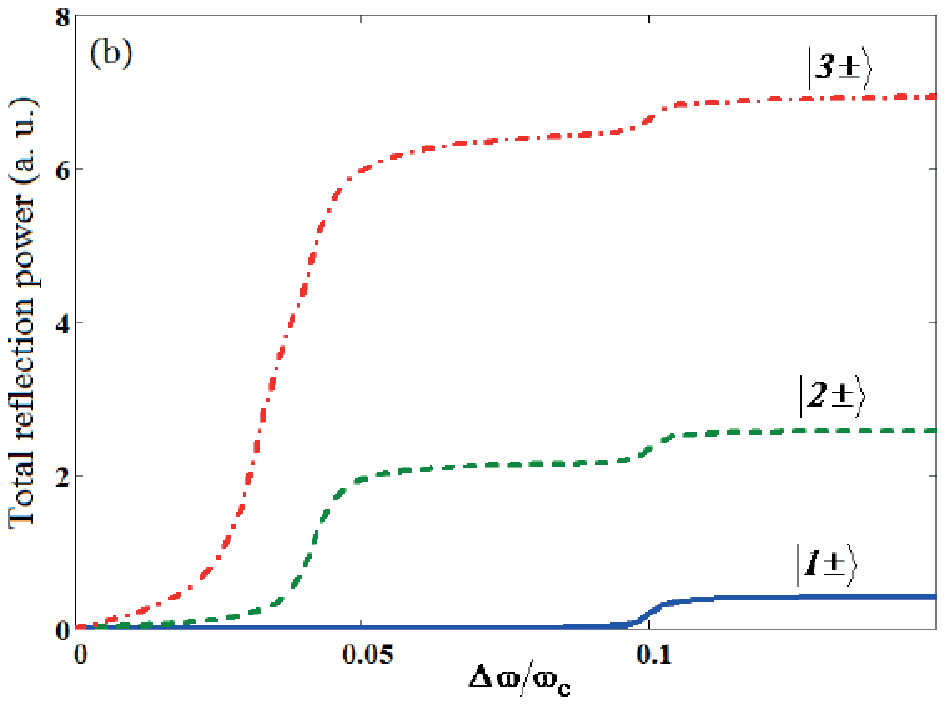}\\
  \caption{Photon blockade staircase. The calculations are made
for $\omega_c=\Omega=3$ GHz, $\lambda=0.1\omega_c$, (a)
$\Gamma=3.33\times 10^{-5}\omega_c=0.1 MHz$, and (b)
$\Gamma=2.66\times 10^{-3}\omega_c=8 MHz$. $|N\pm\rangle$ denotes
the preliminary pumped doublet which successively through one
photon emissions decays to the ground state $|g0\rangle$. The
steps from $|1\pm\rangle$ and $|2\pm\rangle$ doublet ladders are
seen in the $|3\pm\rangle$ doublet ladder. Both plots were
normalized to $10$ MHz. }\label{ladder}
\end{figure}

The application of the prescription encoded in (\ref{RP}) to the
spectrum shown in Fig.\ref{Refl}, provides a typical photon
blockade ladder depicted in Fig.\ref{ladder}. A higher doublet
ladder includes all steps from the lower doublets. The height of a
step scales as the width of the corresponding resonance. The slope
of a step is increased as the emission rate $\Gamma$ of the
photons from the cavity is also increased.

\section{Conclusion}

We develop a theoretical method for the calculation of a microwave
transport in 1D waveguide side coupled to a resonant $N$-photon
cavity with embedded artificial atom (qubit). The method is based
on the projection operator formalism and a non-Hermitian
Hamiltonian approach, which enables us to obtain the analytical
expressions for the probability amplitudes of spontaneous
transitions between the dressed levels of adjacent doublets in
$N$- photon cavity. We show that if the number of the cavity
photons is small the transmitted and reflected spectra reveal a
quadruplet structure with two central peaks and two sidebands. As
the number of the cavity photons is increased the two central
peaks merge giving a classical Mollow triplet.

We considered in detail a single photon transport for the cavity
with two photons. We showed that our theory is in accordance with
the known experiment \cite{Fink08}. Moreover, it predicts the
detection in a single photon experiment of the output photon which
frequency is shifted from that of the input photon by a Rabi
frequency $\Omega_R^{(N-1)}$. We also discussed in detail the
applications of our results to the detection of the photon
blockade ladder which is a direct manifestation of the quantum
nature of light which results from a different space between the
levels in adjacent Rabi doublets.

The results obtained in the paper can be applied to the
investigation of microwave photon transport in superconducting
circuits with embedded superconducting qubits based on Josephson
junctions\cite{You11, Wal04}. The specific properties of the qubit
are encoded in only two parameters: the qubit energy $\Omega$ and
its coupling to the cavity $\lambda$. For example, for a
superconducting qubit $\,\Omega=\sqrt{\varepsilon^2+\Delta^2}$
where $\varepsilon=\frac{2I_q}{\hbar}\left(\Phi_x-\Phi_0/2\right)$
is an external parameter which by virtue of external magnetic
flux, $\Phi_X$ controls the gap between ground and excited states
\cite{Wal00}, $I_q$ is a persistent current along a qubit loop,
$\Phi_0=h/2e$ is a flux quantum. The quantity $\Delta$ is the
qubit's gap at the degeneracy point ($\varepsilon=0$). The
coupling strength $\lambda=g\Delta/\Omega$ \cite{Om10}, where $g$
is the qubit-cavity coupling at the degeneracy point. For a charge
qubit $\,\Omega=\sqrt{\varepsilon^2_J+\varepsilon^2_C}$ where
$\varepsilon_J=2E_J|cos(\pi\Phi_x/\Phi_0|$,
$\varepsilon_C=4E_C(1-2n_g)$, where $E_J$ is a coupling energy of
Josephson junction, $E_C$ is a charging energy, $n_g$ is a
dimensionless gate charge which can be tuned by applying the
voltage $V_g$ to the gate capacitance $C_g$: $n_g=C_gV_g/e$
\cite{Blais04}.

In a more general sense our results can be applied to the
investigation of the photon transport in 1D qubit systems with
small number of cavity photons.

\section*{ACKNOWLEDGMENTS}
The authors are grateful to E. Il'ichev for useful discussions.
This work has been supported by the Russian Science Foundation
under grant No.16-19-10069.
\appendix
\section{The calculation of $H_{XY}$}\label{A1}

With the aid of explicit expressions (\ref{QQ}) and (\ref{PP}) for
$Q$ and $P$ we obtain for the parts $H_{XY}$ of the full
Hamiltonian (\ref{H_ph}) the following expressions:

\begin{multline}\label{HQQ}
    {H_{QQ}} = \frac{1}{2}\hbar \Omega \left| 2 \right\rangle
 \left\langle 2 \right|-\frac{1}{2}\hbar \Omega \left| 1 \right\rangle
    \left\langle 1 \right|
    + \hbar {\omega _c}\left( {N - 1} \right)
    \left| 2 \right\rangle \left\langle 2 \right|\\ + \hbar \lambda
    \sqrt N \left| 1 \right\rangle \left\langle 2 \right| + \hbar
    \lambda \sqrt N \left| 2 \right\rangle \left\langle 1 \right|
      + \hbar {\omega _c}N\left| 1 \right\rangle \left\langle 1
      \right|
\end{multline}

\begin{multline}
{H_{PP}} =  \frac{1}{2}\hbar \Omega \sum\limits_{k} {\left|
{{k_2}} \right\rangle \left\langle {{k_2}} \right|}
 + \hbar {\omega _c}\left( {N - 1} \right)\sum\limits_{k} {\left| {{k_1}} \right\rangle \left\langle {{k_1}}
  \right|} \\
 + \hbar \lambda \sqrt {N - 1} \sum\limits_k {\left| {{k_1}} \right\rangle \left\langle {{k_2}} \right|}
   + \hbar \lambda \sqrt {N - 1} \sum\limits_k {\left| {{k_2}} \right\rangle \left\langle {{k_1}} \right|}\\
 + \sum\limits_k {\hbar {\omega _k}\left| {{k_1}} \right\rangle \left\langle {{k_1}} \right|}
  + \sum\limits_k {\hbar {\omega _k}\left| {{k_2}} \right\rangle \left\langle {{k_2}} \right|}
    \\+ \hbar {\omega _c}\left( {N - 2} \right)\sum\limits_k {\left| {{k_2}} \right\rangle \left\langle {{k_2}}
     \right|}-\frac{1}{2}\hbar \Omega \sum\limits_{k} {\left|
{{k_1}} \right\rangle \left\langle {{k_1}} \right|}
\end{multline}

\begin{equation}\label{HPQ}
    {H_{PQ}} = \hbar \xi \sqrt {N - 1} \sum\limits_k {\left| {{k_2}} \right\rangle \left\langle 2 \right|}
      + \hbar \xi \sqrt N \sum\limits_k {\left| {{k_1}} \right\rangle \left\langle 1 \right|}
\end{equation}

\begin{equation}\label{HQP}
{H_{QP}} = \hbar \xi \sqrt N \sum\limits_k {\left| 1 \right\rangle
\left\langle {{k_1}} \right|}  + \hbar \xi \sqrt {N - 1}
\sum\limits_k {\left| 2 \right\rangle \left\langle {{k_2}}
\right|}
\end{equation}

\section{Calculation of the effective Hamiltonian}\label{A2}

From (\ref{Heff1}) we find the matrix elements of effective
hamiltonian in $Q$ subspace.
\begin{equation}\label{AHeff}
\begin{array}{l}
\left\langle m \right|{H_{eff}}\left| n \right\rangle
 = \left\langle m \right|{H_{QQ}}\left| n \right\rangle \\
 + \sum\limits_{\scriptstyle i,j = 1\hfill\atop
\scriptstyle k,k'\hfill}^2{\left\langle m \right|{H_{QP}}\left|
 {{\varphi _{i,k}}} \right\rangle \left\langle {{\varphi _{i,k}}} \right|
 \frac{1}{{E - {H_{PP}} + i\varepsilon }}\left| {{\varphi_{j,k'}}}
  \right\rangle \left\langle {{\varphi _{j,k'}}} \right|{H_{PQ}}
  \left| n \right\rangle } \\
 = \left\langle m \right|{H_{QQ}}\left| n \right\rangle
  + \sum\limits_{i = 1,k}^{i = 2}\displaystyle {\frac{{\left\langle m \right|{H_{QP}}
  \left| {{\varphi _{i,k}}} \right\rangle \left\langle
  {{\varphi _{{\mathop{\rm i}\nolimits} ,k}}} \right|{H_{PQ}}\left| n
  \right\rangle }}{{E - {E_i}(k) + i\varepsilon }}}
\end{array}
\end{equation}

Fortunately, the matrix elements $\left\langle m
\right|{H_{QP}}\left| {{\varphi _{i,k}}} \right\rangle$ and
$\left\langle {{\varphi _{i,k}}} \right|{H_{PQ}}\left| n
\right\rangle$ do not depend on the photon momentum $k$. The
direct calculations yield:
\begin{equation}\label{AHPQ1}
\left\langle {1|{H_{QP}}|{\varphi _{1,k}}} \right\rangle  =
{a_1}\xi \sqrt N ,\quad\left\langle {2|{H_{QP}}|{\varphi _{1,k}}}
\right\rangle  = {b_1}\xi \sqrt {N - 1}
\end{equation}

\begin{equation}\label{AHPQ2}
\left\langle {1|{H_{QP}}|{\varphi _{2,k}}} \right\rangle  =
{a_2}\xi \sqrt N ,\quad\left\langle {2|{H_{QP}}|{\varphi _{2,k}}}
\right\rangle  = {b_2}\xi \sqrt {N - 1}
\end{equation}

With the use of (\ref{HQQ}) and (\ref{AHPQ1}), (\ref{AHPQ2}) we
obtain for the matrix elements of (\ref{AHeff}):

\begin{subequations}
\begin{equation}\label{AHeff11}
\left\langle {1|{H_{eff}}|1} \right\rangle  = {\omega _C}N -
\frac{1}{2}\Omega  + a_1^2{\xi ^2}N{J_1}(E) + a_2^2{\xi
^2}N{J_2}(E)
\end{equation}

\begin{multline}\label{AHeff22}
\left\langle {2|{H_{eff}}|2} \right\rangle  = {\omega _C}(N - 1) +
\frac{1}{2}\Omega  + b_1^2{\xi ^2}(N - 1){J_1}(E)\\ + b_2^2{\xi
^2}(N - 1){J_2}(E)
\end{multline}

\begin{multline}\label{AHeff12}
\left\langle {1|{H_{eff}}|2} \right\rangle  = \left\langle
{2|{H_{eff}}|1} \right\rangle  = \lambda \sqrt N \\[0.2cm] +
a_1^{}{b_1}{\xi ^2}\sqrt {N(N - 1)} {J_1}(E) + a_2^{}{b_2}{\xi
^2}\sqrt {N(N - 1)} {J_2}(E)
\end{multline}
\end{subequations} where
\begin{equation}\label{AJik}
{J_j}(E) = \sum\limits_k {\frac{1}{{E - {E_j}(k) + i\varepsilon
}}} = \frac{L}{{2\pi }}\int {\frac{{dk}}{{E - {E_j}(k) +
i\varepsilon }}}
\end{equation}

It will be shown below that all quantities $J_j(E)$ in
(\ref{AHeff11}), (\ref{AHeff22}), and (\ref{AHeff12}) are the same
and do not depend on the running energy $E$.
\begin{equation}\label{AJi} {J_j}(E) =  - \frac{{2\pi
i}}{{{{\rm{v}}_g}}}
\end{equation}
where $v_g$ is the  velocity of microwave photons in a waveguide.

Finally, with the use of properties of coefficients $a_i, b_i$
from in (\ref{fi}): $a_1^2+a_2^2=1$, $b_1^2+b_2^2=1$,
$a_1b_1+a_2b_2=0$ we obtain for the matrix elements of $H_{eff}$
the following expressions:

\begin{subequations}
\begin{equation}\label{Heff11}
\left\langle {1|{H_{eff}}|1} \right\rangle  = {\omega _C}N -
\frac{1}{2}\Omega  - jN\Gamma
\end{equation}

\begin{equation}\label{Heff22}
\left\langle {2|{H_{eff}}|2} \right\rangle  = {\omega _C}(N - 1) +
\frac{1}{2}\Omega  - j(N - 1)\Gamma
\end{equation}

\begin{equation}\label{Heff12}
\left\langle {1|{H_{eff}}|2} \right\rangle  = \left\langle
{2|{H_{eff}}|1} \right\rangle  = \lambda \sqrt N
\end{equation}
\end{subequations}

where we introduce the width of the cavity decay rate $\Gamma =
L{\xi ^2}/v_g$.

\section{Calculation of the matrix R}\label{R}

Here we calculate the natrix $R_{m,n}(E)$ which is the matrix
inverse of the matrix $\langle m|(E-H_{eff})|n\rangle$:
\begin{equation}\label{Rmn}
    R_{n,m}(E)=\langle n|\frac{1}{E - H_{eff}}|m\rangle
\end{equation}

From (\ref{Heff11}), (\ref{Heff22}), (\ref{Heff12}) we find the
elements of $R$ matrix (\ref{Rmn}).

\begin{subequations}
\begin{equation}\label{R11}
{R_{11}}(E) = \frac{1}{{D(E)}}\left( {E - {\omega _C}(N - 1) -
\frac{1}{2}\Omega  + j(N - 1)\Gamma } \right)
\end{equation}
\begin{equation}\label{R22}
{R_{22}}(E) = \frac{1}{{D(E)}}\left( {E - {\omega _C}N +
\frac{1}{2}\Omega  + jN\Gamma } \right)
\end{equation}
\begin{equation}\label{R12}
{R_{12}}(E) = {R_{21}}(E) = \frac{{\lambda \sqrt N }}{{D(E)}}
\end{equation}
\end{subequations}
where $D(E)$ is given in (\ref{roots}).

\section{Calculation of transmission matrix (\ref{TM})}\label{A3}

As was shown in Sec.\ref{MT}, $\left\langle {j,k'} \right|T\left|
{i,k} \right\rangle  = \left\langle {{\varphi _j}} \right|T\left|
{{\varphi _i}} \right\rangle  \equiv \xi^2t_{j,i}$. With the aid
of (\ref{AHPQ1}), (\ref{AHPQ2}) we obtain for matrix $t_{ij}$ the
following expressions:

\begin{subequations}
\begin{equation}\label{T11}
\begin{array}{l}
t_{11} = \left( {a_1^2N{R_{11}}({E_1}) + b_1^2(N -
1){R_{22}}({E_1})}
\right.\\[0.2 cm]
{\rm{\quad   \quad\quad\quad\quad\quad\quad }}\left. { +
2{a_1}{b_1}\sqrt {N(N - 1)} {R_{12}}({E_{1}})} \right)
\end{array}
\end{equation}

\begin{equation}\label{T12}
\begin{array}{l}
t_{12}  = \left( {{a_1}{a_2}N{R_{11}}({E_1}) + {b_1}{b_2}(N -
1){R_{22}}({E_1})}
\right.\\[0.2 cm]
\quad\quad \quad \quad \quad \left. { + \sqrt {N(N - 1)}
({a_2}{b_1} + {a_1}{b_2}){R_{12}}({E_1})} \right)
\end{array}
\end{equation}

\begin{equation}\label{T22}
\begin{array}{l}
t_{22}  = \left( {a_2^2N{R_{11}}({E_2}) + b_2^2(N -
1){R_{22}}({E_2})}
\right.\\[0.2 cm]
\quad\quad \quad \quad \quad\left. { + 2{a_2}{b_2}\sqrt {N(N - 1)}
{R_{12}}({E_2})} \right)
\end{array}
\end{equation}

\begin{equation}\label{T21}
\begin{array}{l}
t_{21}  = \left( {{a_1}{a_2}N{R_{11}}({E_2}) + {b_1}{b_2}(N -
1){R_{22}}({E_2})}
 \right.\\[0.2 cm]
\quad\quad \quad \quad \quad\left. { + \sqrt {N(N - 1)}
({a_2}{b_1} + {a_1}{b_2}){R_{21}}({E_2})} \right)
\end{array}
\end{equation}
\end{subequations}

If we substitute in these expressions $a_i, b_i$ for their
explicit forms

\begin{equation}\label{a1b1}
{a_1} = \frac{1}{{\sqrt 2 }}\sqrt {1 - \frac{{\Omega  - {\omega
_c}}}{{{\Omega _R^{(N-1)}}}}}\qquad {b_1} = \frac{1}{{\sqrt 2
}}\sqrt {1 + \frac{{\Omega  - {\omega _c}}}{{{\Omega
_R^{(N-1)}}}}}
\end{equation}

\begin{equation}\label{a2b2}
{a_2} =  - \frac{1}{{\sqrt 2 }}\sqrt {1 + \frac{{\Omega  - {\omega
_c}}}{{{\Omega _R^{(N-1)}}}}} \qquad {b_2} = \frac{1}{{\sqrt 2
}}\sqrt {1 - \frac{{\Omega  - {\omega _c}}}{{{\Omega
_R^{(N-1)}}}}}
\end{equation}

and $R$ from (\ref{R11}), (\ref{R12}), (\ref{R22}), we obtain the
expressions for $t_{ij}$ given in Sec.\ref{MT} in (\ref{t11}),
(\ref{t12}), (\ref{t22}), (\ref{t21}).

\section{Calculation of the photon wavefunction}\label{A4}

As we show in the main text, there are two possible initial states
(\ref{fi}): $|\varphi_{1}\rangle$ and $|\varphi_{2}\rangle$.
Accordingly, there are two wavefunctions (\ref{Ph1}):

\begin{subequations}
\begin{multline}\label{Psi_1}
|{\Psi _1}\rangle  = \left| {{\varphi _{1,k}}} \right\rangle
+ \frac{1}{{{E_1} - {H_{eff}}}}{H_{QP}}\left| {{\varphi _{1,k}}} \right\rangle \\
 + \frac{1}{{{E_1} - {H_{PP}} + i\varepsilon }}{H_{PQ}}\frac{1}{{{E_1} - {H_{eff}}}}{H_{QP}}\left| {{\varphi _{1,k}}} \right\rangle
\end{multline}

\begin{multline}\label{Psi_2}
|{\Psi _2}\rangle  = \left| {{\varphi _{2,k}}} \right\rangle
+ \frac{1}{{{E_2} - {H_{eff}}}}{H_{QP}}\left| {{\varphi _{2,k}}} \right\rangle \\
 + \frac{1}{{{E_2} - {H_{PP}} + i\varepsilon }}{H_{PQ}}\frac{1}{{{E_2} - {H_{eff}}}}{H_{QP}}\left| {{\varphi _{2,k}}} \right\rangle
\end{multline}
\end{subequations}

Next we use the properties of completeness of $P$ and $Q$
($P+Q=1$) and their orthogonality ($PQ=QP=0$) to obtain from
(\ref{Psi_1}) and (\ref{Psi_2})

\begin{equation}\label{APsi1}
\begin{array}{l}
|{\Psi _1}\rangle  = \left| {{\varphi _{1,k}}} \right\rangle \\ +
 \sum\limits_{n,m = 1}^2 {\left| n \right\rangle \left\langle n \right|
 \displaystyle\frac{1}{{{E_1} - {H_{eff}}}}\left| m \right\rangle \left\langle m
 \right|{H_{QP}}\left| {{\varphi _{1,k}}} \right\rangle } \\
 + \sum\limits_{\scriptstyle i,j = 1\hfill\atop
\scriptstyle k,k'\hfill}^2 {\left\{ {\left| {{\varphi _{i,k}}}
\right\rangle \left\langle {{\varphi _{i,k}}} \right|
\displaystyle\frac{1}{{{E_1} - {H_{PP}} + i\varepsilon }}\left|
{{\varphi _{i,k'}}}
 \right\rangle } \right.} \\
\left. { \times \left\langle {{\varphi _{j,k'}}}
\right|{H_{PQ}}\left| n \right\rangle \left\langle n
\right|\displaystyle\frac{1}{{{E_1} - {H_{eff}}}}\left| m
\right\rangle \left\langle m \right|{H_{QP}}\left| {{\varphi
_{1,k}}} \right\rangle } \right\}
\end{array}
\end{equation}

\begin{equation}\label{APsi2}
\begin{array}{l}
|{\Psi _2}\rangle  = \left| {{\varphi _{2,k}}} \right\rangle\\  +
 \sum\limits_{n,m = 1}^2 {\left| n \right\rangle \left\langle n \right|
 \displaystyle\frac{1}{{{E_2} - {H_{eff}}}}\left| m \right\rangle \left\langle m
 \right|{H_{QP}}\left| {{\varphi _{2,k}}} \right\rangle } \\
 + \sum\limits_{\scriptstyle i,j = 1\hfill\atop
\scriptstyle k,k'\hfill}^2 {\left\{ {\left| {{\varphi _{i,k}}}
\right\rangle \left\langle {{\varphi _{i,k}}} \right|
\displaystyle\frac{1}{{{E_2} - {H_{PP}} + i\varepsilon }}\left|
{{\varphi _{i,k'}}}
 \right\rangle } \right.} \\
\left. { \times \left\langle {{\varphi _{j,k'}}}
\right|{H_{PQ}}\left| n \right\rangle \left\langle n
\right|\displaystyle\frac{1}{{{E_2} - {H_{eff}}}}\left| m
\right\rangle \left\langle m \right|{H_{QP}}\left| {{\varphi
_{2,k}}} \right\rangle } \right\}
\end{array}
\end{equation}

From these equations it follows immediately the expressions
(\ref{Psi_1a}) and (\ref{Psi_2a}), which we write here in the
following form:
\begin{multline}\label{Apsi1}
    |{\Psi _1}\rangle  = \left| {{\varphi _{1,k}}} \right\rangle  +
\sum\limits_{m,n} {\left| n \right\rangle }
{R_{nm}}({E_1})\left\langle m \right|{H_{QP}}\left| {{\varphi
_{1,k}}} \right\rangle \\ + {\xi ^2}\sum\limits_{q,i}
{\frac{{\left| {{\varphi _{i,q}}} \right\rangle
{t_{i1}}}}{{{E_1}(k) - {E_i}(q) + i\varepsilon }}}
\end{multline}

\begin{multline}\label{Apsi2}
    |{\Psi _2}\rangle  = \left| {{\varphi _{2,k}}} \right\rangle  +
\sum\limits_{m,n} {\left| n \right\rangle }
{R_{nm}}({E_2})\left\langle m \right|{H_{QP}}\left| {{\varphi
_{2,k}}} \right\rangle \\ + {\xi ^2}\sum\limits_{q,i}
{\frac{{\left| {{\varphi _{i,q}}} \right\rangle
{t_{i2}}}}{{{E_2}(k) - {E_i}(q) + i\varepsilon }}}
\end{multline}

In order to obtain photon wavefunction in a configuration space we
multiply (\ref{Apsi1}) and (\ref{Apsi2}) from the left by bra
vector $\langle x|$, and taking into account that $\langle
x|n\rangle=0$, $\langle
x|\varphi_{i,k}\rangle=e^{ikx}|\varphi_{i}\rangle$, we obtain:

\begin{equation}\label{APsi1x}
\left\langle {x|{\Psi _1}} \right\rangle  = {e^{ikx}}\left|
{{\varphi _1}} \right\rangle  + {\xi ^2}\sum\limits_{i = 1}^2
{{J_{i,1}}{t_{i1}}\left| {{\varphi _i}} \right\rangle }
\end{equation}

\begin{equation}\label{APsi2x}
\left\langle {x|{\Psi _2}} \right\rangle  = {e^{ikx}}\left|
{{\varphi _2}} \right\rangle  + {\xi ^2}\sum\limits_{i = 1}^2
{{J_{i,2}}{t_{i2}}\left| {{\varphi _i}} \right\rangle }
\end{equation}

where
\begin{equation}\label{AJ1}
{J_{i,j}} = \sum\limits_q {\frac{{{e^{iqx}}}}{{{E_j}(k) - {E_i}(q)
+ i\varepsilon }}}
\end{equation}

Below we calculate the quantities $J_{i,j}$. The result is as
follows:
\begin{equation}\label{AJ11}
    J_{11}=J_{22}=-i\frac{L}{v_g}e^{ik|x|}
\end{equation}
\begin{equation}\label{AJ12}
    J_{12}=-i\frac{L}{v_g}e^{i(k-k_R)|x|}
\end{equation}

\begin{equation}\label{AJ21}
    J_{21}=-i\frac{L}{v_g}e^{i(k+k_R)|x|}
\end{equation}
where $k_R=\Omega_R^{(N-1)}/v_g$.

 With the account of these results we obtain for the photon
wavefunctions (\ref{APsi1x}), (\ref{APsi2x}) the expressions
(\ref{Phwf1}) and \ref{Phwf2}) from the main text.

\section{Calculation of $J_{i,j}$}

From (\ref{En}) we find the energy difference in the denominator
of (\ref{AJ1}):
\begin{equation}\label{endif}
\begin{array}{l}
{E_i}(k) - {E_i}(q) = {\omega _k} - {\omega _q} = {v_g}(k -
q);\\[0.2cm]
{E_1}(k) - {E_2}(q) = {\omega _k} - {\omega _q} + {\Omega
_R^{(N-1)}} =
{v_g}(k - q + k_R);\\[0.2cm]
{E_2}(k) - {E_1}(q) = {\omega _k} - {\omega _q} - {\Omega _R} =
{v_g}(k - q - k_R)
\end{array}
\end{equation}

As an example we calculate below the quantity $J_{12}$
(\ref{AJ12}) where we substitute the summation over $q$ for the
integration:
\begin{equation}\label{AJ12a}
{J_{12}} = \frac{L}{{2\pi }}\int\limits_{ - \infty }^{ + \infty }
{\frac{{{e^{iqx}}}}{{{\omega _k} - {\omega _q} - \Omega_R^{(N-1)}
+ i\varepsilon }}} dq
\end{equation}

The main contribution to this integral comes from the region where
${\omega _q} \approx {\omega _k}-\Omega_R^{(N-1)}$. Since
$\omega_q$ is the even function of $q$, it can be approximated
away from the cutoff frequency as $\omega_q\equiv v_g |q|$. In
this case the poles of the integrand (\ref{AJ12a}) in the $q$
plane are located near the points $q\approx\pm q_0$ where
$q_0=(k-k_R)$. From denominator in (\ref{AJ12a}) we see that one
pole is located in the upper half of the $q$ plane,
$q=q_0+i\varepsilon$, the other pole is located in the lower half
of the $q$ plane, $q=-q_0-i\varepsilon$. For positive $x$, when
calculating the integral (\ref{AJ12a}) we must close the path in
the upper plane. For negative $x$ the path should be closed in
lower plane. Thus, we obtain:

\begin{equation}\label{A0}
J_{12} =  - i\frac{{L}}{\hbar v_g}{e^{i(k-k_R)|x|}}
\end{equation}
The quantities $J_{11}$, $J_{22}$ (\ref{AJ11}) and $J_{21}$
(\ref{AJ21}) can be calculated by the same procedure.

\end{document}